%% file: ANA-STDM-2022-16-PAPER.tex
\documentclass[PAPER, atlasdraft=false,,UKenglish,cernpreprint, texmf, orcidlogo]{atlasdoc}
\usepackage{atlaspackage}
\usepackage{float}
\usepackage{etoolbox}
\usepackage{mathrsfs}
\usepackage{footnote}
\usepackage{multirow}
\usepackage{subfig}
\usepackage{amsmath}
\usepackage{graphicx}
\usepackage{array}
\usepackage{comment}
\usepackage{tabularx}
\usepackage{booktabs}
\usepackage{pifont}
\usepackage{tikz}
\usepackage{tikz-feynman}
\usepackage{feynmp}
\usepackage{xcolor}
\usetikzlibrary{matrix,calc}
\usetikzlibrary{arrows.meta}
\usepackage{makecell}
\usepackage{caption}
\usepackage{cleveref}

\usepackage{atlasbiblatex}

\usepackage[hepprocess]{atlasphysics}
\usepackage{atlascontribute}
\usepackage{hyperref}

\graphicspath{{logos/}{figures/}}

\usepackage{ANA-STDM-2022-16-PAPER-defs}
\AtlasTitle{Observation of the electroweak production of $\gamma\gamma jj$ at $\sqrt{s}=13$ TeV in 140 fb$^{-1}$ of $pp$ collision data with the ATLAS detector}

\AtlasAbstract{%
This Letter presents the observation of the electroweak production of two isolated photons in association with two jets together with measurements of the corresponding fiducial and differential cross sections. The analysis is performed using proton--proton collision data collected by ATLAS at the LHC at $\sqrt{s}$ = 13 TeV and with an integrated luminosity of 140 fb$^{-1}$. The electroweak production of two isolated photons in association with two jets is observed with a significance of $6.2 \sigma $ and with a fiducial cross-section of  $\sigma^{\mathrm{EW}}_{\mathrm{obs}}= 13.8^{+3.0}_{-2.6}\, \mathrm{fb}$, in agreement with the Standard Model prediction of $\sigma^{\mathrm{EW}}_{\mathrm{obs}}= 17.1^{+2.4}_{-2.4}\, \mathrm{fb}$.
Differential cross-sections are measured as a function of various kinematical variables probing photon and jet properties and compared with predictions.
In addition, fiducial and differential cross-sections are reported for the combined QCD and electroweak production of two isolated photons and two jets. The measurement is also interpreted in the context of an effective field theory, and  limits are set on the Wilson coefficients of several dimension-8 operators parameterising anomalous quartic gauge couplings.
}

\usepackage{authblk}
\makeatletter
\renewcommand\AB@affilsepx{, \protect\Affilfont}
\makeatother
\renewcommand\Affilfont{\itshape\small} %

\AtlasRefCode{STDM-2022-16}

\PreprintIdNumber{CERN-EP-2026-219}

\AtlasJournal{Phys.\ Lett.\ B.}
\AtlasCoverEgroupAnalysisTeam{atlas-stdm-2022-16-analysis-team@cern.ch}

\hypersetup{pdftitle={ATLAS document},pdfauthor={The ATLAS Collaboration}}

\begin{document}

\maketitle

%

%
%

\section{Introduction}

The investigation of the electroweak (EW) production of two vector bosons accompanied by two jets offers a strong test of the Standard Model (SM), as this process is sensitive to gauge-boson self-interactions.  The study of vector-boson scattering (VBS) phenomena ($VV\rightarrow VV$, where $V=W,Z$), provides direct access to the EW symmetry-breaking mechanism within the SM and allows constraints to be placed on anomalous gauge-boson couplings. Stronger limits extend the reach to new-physics energy scales in the multi-\TeV domain and create opportunities to identify potential signs of physics beyond the SM through a model-independent approach, such as effective field theory (EFT). In recent years  the ATLAS and CMS experiments at the LHC~\cite{Evans:2008zzb} have observed and measured several EW processes involving gauge bosons~\cite{STDM-2019-27,STDM-2019-28,STDM-2020-16,STDM-2018-31,STDM-2018-36,STDM-2022-06,CMS-SMP-22-001,CMS-SMP-21-005,CMS-SMP-20-014,CMS-SMP-19-005,CMS-SMP-21-011,CMS-SMP-21-001}. However, one important process that remains to be studied is the EW production of two photons in association with two jets (EW-\yyjj).
This process  is sensitive to anomalous gauge-boson interactions~\cite{Chapon:2009hh,Eboli:2000ad} through the $VV\gamma\gamma$ quartic gauge coupling.
In addition, EW-\yyjj events constitute one of the main background sources in $H\rightarrow\gamma\gamma$ measurements in the vector-boson fusion (VBF) production mode~\cite{HIGG-2020-08,HIGG-2020-16} and therefore an accurate modelling of the EW-\yyjj production can help reduce uncertainties associated with the estimate of this background.

\begin{figure}[!htbp]
\centering
\subfloat[]{\includegraphics[width=0.24\textwidth]{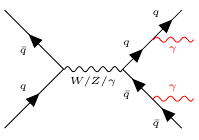}
\label{fig:EW_a}
}
\subfloat[]{\includegraphics[width=0.24\textwidth]{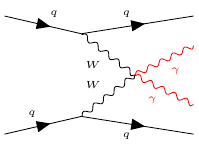}
}
\subfloat[]{\includegraphics[width=0.24\textwidth]{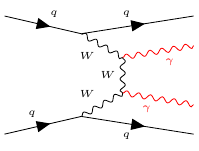}
\label{fig:EW_c}
}\\
\subfloat[]{\includegraphics[width=0.24\textwidth]{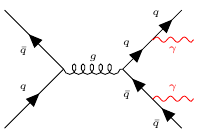}
\label{fig:QCD_a}
}
\subfloat[]{\includegraphics[width=0.24\textwidth]{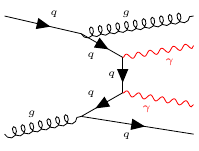}
\label{fig:QCD_b}
}
\subfloat[]{\includegraphics[width=0.24\textwidth]{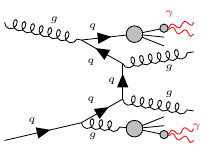}
\label{fig:yj}
}
\caption{Representative Feynman diagrams for the (a, b, c) EW-\yyjj production, (d, e) QCD-\yyjj production and (f) non-prompt photon production in association with jets. The grey circles indicate hadronisation and hadron decays.  Only (b, c) show a vector boson scattering topology. }

\label{fig:Diagram}
\end{figure}

This Letter presents the observation of the EW production of two isolated photons in association with two jets in $pp$ collisions at $\sqrt{s}=13~\TeV$ using the data collected by ATLAS during Run 2. Fiducial and differential cross-sections are measured for the EW-\yyjj production.
The differential cross-sections are measured for observables sensitive to the properties of photons and jets, the VBS topology, and potential beyond the Standard Model (BSM) effects.
All the observables are defined in Section~\ref{sec:DiffXS}.
In proton--proton ($pp$) collisions, two main types of processes lead to the \yyjj final state.
The first category corresponds to purely EW production at Born level, of order
$\mathcal{O}(\alpha_{\mathrm{EW}}^{4})$, where $\alpha_{\mathrm{EW}}$ denotes the EW coupling constant.
This contribution includes VBS topology and other EW diagrams leading to the same final state and is considered as the signal.
The second category includes processes mediated at Born level by both QCD and EW interactions of order $\mathcal{O}(\alpha_{\mathrm{EW}}^{2}\alpha_{\mathrm{s}}^{2})$, where $\alpha_{\textrm{s}} $ is the strong interaction coupling constant.  This contribution is referred to as QCD-\yyjj and is considered as background.
Fiducial and differential cross-sections are also reported for the combined QCD+EW-\yyjj production. Representative Feynman diagrams for EW- and QCD-induced $\gamma\gamma jj$ production are shown in
Figures~\ref{fig:EW_a}--\ref{fig:EW_c} and~\cref{fig:QCD_a,fig:QCD_b}, respectively.
In both production categories, \textit{prompt} photons are defined as photons produced either directly in
the hard-scattering process or via single- or double-fragmentation of high-$\pT$ partons.

The contributions from \textit{non-prompt}  photons is dominated by QCD multijet events in which a pair of photon candidates originates from hadron decays, such as $\pi^0\rightarrow\gamma\gamma $ (see Figure~\ref{fig:Diagram}\subref{fig:yj}). These events are treated as background and are suppressed by selecting events with photons isolated from the hadronic activity.  The remaining non-prompt backgrounds are estimated with a data-driven technique, following closely the procedure in Ref.~\cite{STDM-2017-30}.
Events from the production of Higgs bosons ($H\rightarrow\gamma\gamma $) are considered as background, since they occur at a different $\alpha_{\textrm{EW}}$ order than the QCD-\yyjj.

This Letter is structured as follows: the ATLAS detector is presented in Section~\ref{sec:ATLAS}, while the data and the simulated samples used in the analysis are described in Section~\ref{sec:Samples}. Object reconstruction and event selection are described in Section~\ref{sec:ObjEvntSel}.  The background estimate is presented in Section~\ref{sec:background} and the fiducial phase space and the unfolding technique are described in Sections~\ref{sec:IncXS} and~\ref{sec:DiffXS} respectively. Systematic uncertainties and their impact on the analysis are described in Section~\ref{sec:uncertainties}. Finally, results are discussed in Section~\ref{sec:results} and interpreted in the context of EFT in Section~\ref{sec:EFT}. Section~\ref{sec:conclusion} shows the conclusions.


\section{ATLAS detector}
\label{sec:ATLAS}
\newcommand{\AtlasCoordFootnote}{%
ATLAS uses a right-handed coordinate system with its origin at the nominal interaction point (IP)
in the centre of the detector and the \(z\)-axis along the beam pipe.
The \(x\)-axis points from the IP to the centre of the LHC ring,
and the \(y\)-axis points upwards.
Polar coordinates \((r,\phi)\) are used in the transverse plane,
\(\phi\) being the azimuthal angle around the \(z\)-axis.
The pseudorapidity is defined in terms of the polar angle \(\theta\) as \(\eta = -\ln \tan(\theta/2)\) and is equal to the rapidity
$ y = \frac{1}{2} \ln \left( \frac{E + p_z}{E - p_z} \right) $ in the relativistic limit.
Angular distance is measured in units of \(\Delta R \equiv \sqrt{(\Delta y)^{2} + (\Delta\phi)^{2}}\).}

The ATLAS experiment~\cite{PERF-2007-01} at the LHC is a multipurpose particle detector
with a forward--backward symmetric cylindrical geometry and a near \(4\pi\) coverage in
solid angle.\footnote{\AtlasCoordFootnote}
It consists of an inner tracking detector surrounded by a thin superconducting solenoid
providing a \qty{2}{\tesla} axial magnetic field, electromagnetic (EM) and hadronic calorimeters, and a muon spectrometer.
The inner tracking detector covers the pseudorapidity range \(|\eta| < 2.5\).
It consists of silicon pixel, silicon microstrip, and transition radiation tracking detectors.
Lead/liquid-argon (LAr) sampling calorimeters provide EM energy measurements
with high granularity within the region \(|\eta|< 3.2\).
A steel/scintillator-tile hadronic calorimeter covers the central pseudorapidity range (\(|\eta| < 1.7\)).
The endcap and forward regions are instrumented with LAr calorimeters
for EM and hadronic energy measurements up to \(|\eta| = 4.9\).
The muon spectrometer surrounds the calorimeters and is based on
three large superconducting air-core toroidal magnets with eight coils each.
The field integral of the toroids ranges between \num{2.0} and \qty{6.0}{\tesla\metre}
across most of the detector.
The muon spectrometer includes a system of precision tracking chambers up to \(|\eta| = 2.7\) and fast detectors for triggering up to \(|\eta| = 2.4\).
The luminosity is measured mainly by the LUCID--2~\cite{LUCID2} detector, located close to the beampipe.
A two-level trigger system was used to select events~\cite{TRIG-2016-01}.
The first-level trigger is implemented in hardware and uses a subset of the detector information
to accept events at a rate close to \qty{100}{\kHz}.
This is followed by a software-based trigger that
reduced the accepted rate of complete events to \qty{1.25}{\kHz} on average
depending on the data-taking conditions.
A software suite~\cite{SOFT-2022-02} is used in data simulation, in the reconstruction
and analysis of real and simulated data, in detector operations, and in the trigger and data acquisition
systems of the experiment.


%
%

\section{Data and simulated event samples}
\label{sec:Samples}

This analysis is based on the full \(pp\) collisions data sample collected by the ATLAS detector during the Run-2 data-taking period of the LHC (2015--2018), at a centre-of-mass energy of $\sqrt{s} =13~\TeV$ with a total integrated luminosity of 140.1$\pm$1.2~\ifb~\cite{DAPR-2021-01}.
Events are required to satisfy quality requirements ensuring that the detector was in a stable operating condition~\cite{DAPR-2018-01}.

Monte Carlo (MC) simulations are used to model the signal, the background processes and the detector response. All MC samples are processed through the full ATLAS detector simulation~\cite{SOFT-2010-01} based on \GEANT~\cite{Agostinelli:2002hh} and reconstructed with the same software as the data.
The effect of multiple interactions in the same or nearby bunch crossing (pile-up) was simulated by generating a set of inelastic proton–proton collision events using  \PYTHIA[8.1]~\cite{Sjostrand:2007gs} with the \NNPDF[2.3lo]~\cite{Ball:2012cx}  set of parton distribution functions (PDF) and the A3 set of tuned parameters~\cite{ATL-PHYS-PUB-2016-017}.  These events are then overlaid on top of the original hard-scatter signal and background events.

The EW-$\gamma\gamma jj$ signal process is simulated at leading-order (LO) accuracy ($\mathcal{O}(\alpha_{\textrm{EW}}^4)$) using \MGNLO[2.6.7]~\cite{Alwall:2014hca} with the default renormalisation and factorisation scale choice~\cite{Alwall:2014hca}.
The sample utilises the \NNPDF[3.1nlo]~\cite{NNPDF:2017mvq} PDF set.
\MGNLO is interfaced with \PYTHIA[8.244]~\cite{Sjostrand:2014zea} for parton showering (PS) and hadronisation. The A14 PS tune~\cite{ATL-PHYS-PUB-2014-021} is employed with the \texttt{dipoleRecoil} option enabled for a better treatment of the VBS processes~\cite{Jager:2020hkz,ATL-PHYS-PUB-2019-004}. Photon isolation criteria are applied at the parton level using the smooth-cone isolation prescription described in Ref.~\cite{Frixione:1998jh}. This method ensures the removal of contributions from photons originating from quark or gluon fragmentation in an infrared-safe manner, valid to all orders of perturbation theory. %
An alternative EW-$\gamma\gamma jj$ sample is generated with \SHERPA[2.2.12]~\cite{Bothmann:2019yzt} using
Matrix Element (ME) calculation at LO accuracy in $\alphas$ with up to one additional parton beyond the two leading jets.
To ensure a smooth transition to the PS and combine different jet multiplicities, ME are matched and merged with the \SHERPA PS—based on Catani–Seymour dipole subtraction~\cite{Schumann:2007mg}—using the MEPS@LO prescription~\cite{Schalicke:2005nv}. The \NNPDF[3.0nnlo]~\cite{Ball:2014uwa} PDF set is used, and the smooth-cone isolation prescription is applied with the same parameters as described above.

The QCD-$\gamma\gamma jj$ production, where jets originate from strong interaction, is modelled using \SHERPA[2.2.4]~\cite{Bothmann:2019yzt}. In the ME calculation, the diphoton system is accompanied by up to one additional parton at next-to-leading-order (NLO) in QCD, while configurations with up to three additional partons beyond the first are at LO accuracy. The \MEPSatNLO~\cite{Hoeche:2011fd,Hoeche:2012yf,Catani:2001cc,Hoeche:2009rj} prescription is employed for merging of various jet multiplicities. The virtual QCD corrections are provided by the \OPENLOOPS library~\cite{Cascioli:2011va,Denner:2016kdg}.  The \NNPDF[3.0nnlo] PDF set is used in the simulation. In addition, the loop-induced $gg \to \gamma\gamma$ box process is included at LO accuracy.
Alternative QCD-$\gamma\gamma jj$ samples are generated with \MGNLO[2.3.3]~\cite{Alwall:2014hca} with up to two partons at NLO accuracy in the ME calculation. The showering and subsequent hadronisation is performed using \PYTHIA[8.210] with the A14 tune, using the \NNPDF[2.3lo]  PDF set. The different jet multiplicities are merged using the \FXFX prescription~\cite{Frederix:2012ps} implemented in \MGNLO[2.3.3].  The PDF input set in use for event generation is the \NNPDF[3.0nlo] PDF set.
The parameterisation of the effects of EFT operators on the EW-\yyjj signal is derived using MC samples generated with \SHERPA[3]~\cite{Sherpa:2024mfk} in a merged set-up with LO ME including up to one additional jet beyond the two leading jets, employing the same generator settings as those used for the alternative EW-\yyjj sample described above and the \NNPDF[3.0nnlo] PDF set.
The $Z/\gamma^*$  Drell–Yan (DY) and $W$ to electrons process are modelled using \POWHEGBOX[v2]~\cite{Alioli:2010xd,Nason:2004rx} using the \CT[10]~\cite{Lai:2010vv} PDF set and interfaced with \PYTHIA[8.186] for PS and hadronisation.  The VBF $Z/\gamma^*$ and $W$ processes are simulated using \SHERPA[2.2.1] using the \NNPDF[3.0nnlo] PDF set, and including up to four additional jets at LO in the ME calculation. Events from the production of Higgs bosons were simulated using \POWHEGBOX[v2]~\cite{Hamilton:2013fea,Hamilton:2015nsa,Frixione:2007vw} with the \PDFforLHC[15nnlo] PDF set~\cite{Butterworth:2015oua} for the gluon--gluon fusion, vector-boson
fusion, and associated production with a vector-boson, while the \ttH production mode used the \NNPDF[3.0nlo] PDF set.
The production of \tHq events was modelled using \MGNLO[2.6.0] with the \NNPDF[3.0nlo] PDF. For all Higgs-boson production modes, the simulations were interfaced to \PYTHIA[8] for PS and non-perturbative effects, with parameters set according to the \AZNLO tune~\cite{STDM-2012-23}, except for the \ttH and \tHq samples, which used the A14 tune~\cite{ATL-PHYS-PUB-2014-021}.


\section{Object reconstruction and event selection}
\label{sec:ObjEvntSel}

This analysis targets a final state with two photons and two jets in a VBS topology, characterised by two jets produced in opposite hemispheres of the detector and with a large dijet invariant mass.  Photons are reconstructed by clustering EM calorimeter cells with significant energy deposits into \textit{topological clusters}  based on the significance of energy deposits~\cite{PERF-2014-07}, and calibrated following the procedure described in Ref.~\cite{EGAM-2021-02}. Photons are also required to satisfy the cut-based \textit{tight} identification criteria, based on shape observables of the EM shower, as described in Ref.~\cite{EGAM-2021-02}.
Events are collected with the lowest unprescaled di-photon trigger~\cite{TRIG-2018-05}, with a trigger threshold of 35 (25)~\GeV for the leading (subleading) photon candidate, ordered by \pt. The leading (subleading) photon candidate is required to have a transverse momentum greater than 40 (30)~\GeV, ensuring that photons are in the trigger efficiency plateau.
Photon candidates are also required to satisfy $|\eta|<2.37$, excluding the EM calorimeter barrel--endcap transition region ($1.37<|\eta|<1.52$).
The momenta of the photon candidates are corrected for the position of their associated reconstructed primary interaction vertex.  The primary vertex is selected using a neural-network algorithm following the procedure described in Ref.~\cite{HIGG-2013-08}. At least one primary vertex is required per event with at least two associated tracks with $\pT>0.5~\GeV$.
Photon isolation is defined following the procedure in Ref.~\cite{STDM-2017-30} based on the transverse isolation energy, $\Eiso$, measured within an isolation cone of size $\Delta R = 0.2$. The energy deposited by the photon itself, is excluded from the isolation energy. A correction is then applied for the residual leakage of the photon shower into the isolation region, as well as for contributions from pile-up and the underlying event, following the procedures described in Ref.~\cite{PERF-2014-07} and Ref.~\cite{Cacciari:2007fd}.
Isolation criteria are then obtained  by requiring $\Eiso $ to be lower than 5\% of the transverse energy of the photon itself. Photons are also required to have an angular separation of at least $\Delta R>0.4 $ to avoid overlap and to minimise the correlation between the \Eiso of the two photons.

Jets are reconstructed using the \antikt algorithm~\cite{Cacciari:2008gp,Cacciari:2006sm,Fastjet} with a radius parameter of $R=0.4$ and using a particle-flow~\cite{PERF-2015-09} procedure that combines calorimeter deposits and tracks to determine jet properties.  Jets are calibrated using in situ measurements of the jet energy scale and MC based methods, following the procedure described in Ref.~\cite{JETM-2018-05}.  Jets are required to have a transverse momentum of $\pT>25~\GeV $ and pseudorapidity $|\eta|<4.4$. Jets with $\pT < 60~\GeV $ and $|\eta|<2.4$ are discarded if they are classified as pile-up jets based on the jet vertex tagger (JVT)~\cite{PERF-2014-03}.

The two jets of highest \pT are required to have $\pT>60~\GeV$ to enhance the EW-\yyjj production mode. All jets and photons are required to have an angular separation of at least $\Delta R(j,\gamma)>0.4 $ to avoid ambiguities.  %
Events are selected if each of the two leading photons is separated from each of the two leading jets by at least $\Delta R=0.6 $.
The inclusive VBS region is defined by requiring the two jets to have separation in rapidity of at least $|\Delta y_{jj}|>2.0 $ and to be in opposite hemispheres of the detector: $y_{j_1}\cdot y_{j_2}<0$. Additionally, the invariant mass of the two jets, \mjj, has to be greater than 500~\GeV.
Jets in the rapidity interval between the two leading jets are referred to as \textit{gap jets}.
The VBS inclusive region is then divided into a Signal Region (EW-SR) and three Control Regions (QCD-CR1, QCD-CR2, and QCD-CR3) based on the number of gap-jets and the centrality of the photons relative to the two leading jets, \xiyy, defined as:

\begin{equation}
\xiyy=\left|\frac{y_{\gamma_1\gamma_2}-\frac{y_{j_1}+y_{j_2}}{2}}{y_{j_1}-y_{j_2}}\right|,
\end{equation}
where $y_{\gamma_1\gamma_2}$ is the rapidity of the di-photon system.
Table~\ref{tab:baselineselection} shows the object and event selection criteria as well as the definition of the EW-SR and of the three Control Regions.

\begin{table}[h]
\caption{Overview of the object and event selection for the VBS region at both detector- and particle-level and definition of the signal and control regions based on the di-photon centrality, \xiyy, and  the number of gap-jets, \Ngy.}
\renewcommand{\arraystretch}{1.2} %
\centering
\begin{tabular}{lc}
\toprule
& \textbf{Event selection}  \\
\midrule

Photons &    $\pTyo>40$ \GeV{}, $\pTyt>30$ \GeV{}  \\
&  $|\eta_{\gamma_{1(2)}}|<2.37$ (excl. $1.37<|\eta_{\gamma_{1(2)}}|<1.52$) \\
& both Tight identification  \\
& $E^{\mathrm{iso}}_{\mathrm{T}, \gamma_{1(2)}} < 0.05 \cdot E_{\mathrm{T}, \gamma_{1(2)}}$ (detector- level) /  $E^{\mathrm{iso}}_{\mathrm{T}, \gamma_{1(2)}} < 0.09 \cdot E_{\mathrm{T}, \gamma_{1(2)}}$ (particle-level)\\
&  $\Delta R_{\gamma\gamma}>0.4$ \\
\midrule
Jets  &  $p_{\mathrm{T}, j}>25$ \GeV{} \\
&  $|\eta_{j}|<4.4$ \\
&  $N_{j}\geq 2$ \\
&  $p_{\mathrm{T}, j_{1(2)}}>60$ \GeV{} \\
\midrule
VBS selection   & $|\Delta y_{jj}|>2.0$ \\
& $y_{j_1} \cdot y_{j_2}<0$ \\
& $m_{jj}>500$~\GeV \\

&            $\Delta R(\gamma_{i},j_{k})>0.6\,(i,k=1,2) $
\\
\midrule

Region   &  \\
\midrule
EW-SR  &  $ \xiyy<0.5$, $\Ngy=0$  \\
QCD-CR1  &  $ \xiyy>0.5$, $\Ngy=0$  \\
QCD-CR2 &  $ \xiyy<0.5$, $\Ngy=1$  \\
QCD-CR3 &  $ \xiyy>0.5$, $\Ngy=1$  \\
\bottomrule
\end{tabular}
\label{tab:baselineselection}
\end{table}

The fiducial and differential cross-sections are measured in a phase space defined at particle-level, as shown in Table~\ref{tab:baselineselection}. At particle-level only photons not originating from hadron decays are considered as prompt photons, and the two photons with the highest transverse momentum are selected. Contributions from the underlying event such as multi-parton interactions and beam remnants are removed from the particle-level isolation energy $E_{\mathrm{T},\gamma_{1(2)}}^{\mathrm{iso}}$ to reproduce the subtraction applied also at detector-level, as in Ref.~\cite{STDM-2017-30}. The isolation requirement at particle level is obtained by requiring $E_{\mathrm{T},\gamma_{1(2)}}^{\mathrm{iso}}$ to be less than 9\% of the photon transverse energy, instead of 5\% as at detector-level. This difference accounts for the non-compensating nature of the ATLAS calorimeter and was optimised to reduce the number of photons satisfying the isolation requirement at particle(detector)-level but not at detector(particle)-level.

Particle-level jets are built using the anti-$k_t$ algorithm with radius $R=0.4$ from stable particles, excluding muons, neutrinos, and particles originating from pile-up vertices. Jets are discarded if they are within $\Delta R=0.4$ of a photon or each other. The VBS enriched region is defined as in the detector-level case.


\section{Background estimate}
\label{sec:background}
The dominant background in the EW-$\gamma\gamma jj$ production is the irreducible contribution from QCD production of  $\gamma\gamma jj$. Differences between the QCD-\yyjj simulation and data are observed in both the overall normalisation and the shapes of certain observables, most notably $m_{jj}$. To mitigate the limitations of MC modelling, the QCD-\yyjj background is estimated by using a partially data-driven approach, with QCD-CR used to extrapolate into the EW-SR. The strategy is detailed in Section~\ref{sec:IncXS}.

The second largest background arises from non-prompt photons, predominantly from neutral pions decaying into photon pairs. It is estimated with the data-driven two-dimensional sideband method described in Ref.~\cite{STDM-2017-30}. The method extrapolates the non-prompt background from dedicated control regions to the prompt-photon signal region using a simultaneous likelihood fit across 16 regions, defined according to whether each photon candidate passes or fails the identification and isolation requirements. The region in which both photon candidates satisfy both requirements is defined as the prompt-photon signal region, while the remaining 15 regions serve as control regions.
The likelihood model is constructed using probability density functions $f_{p,i}$, which parameterize the probability for an event from process pp $p$ ($\gamma\gamma$, $j\gamma$, $\gamma j$, $jj$,  electrons misidentified as photons, or pile-up $\gamma \gamma$) to populate each of the 16 regions ($i=1,...16 $).The region with $i=1$ is the prompt-photon signal region.
Following Ref.~\cite{STDM-2017-30}, each $f_{p,i}$ is factorised into the identification and isolation efficiencies of the leading ($n=1$) and sub-leading ($n=2$) photon candidates, $\varepsilon^{\mathrm{id(iso)}}_{p,n}$, together with the correlation factors between the isolation and identification of the same photon $n$ ($R^{\text{iso-id}}_{p,n}$ ), and the correlation between the identification (isolation) of the two photon candidates ($R^{\text{id(iso)}}_{p}$).

Most parameters associated with the non-prompt components are left free in the fit because their normalisations are not reliably predicted by simulation. The correlation factors are taken from simulation or dedicated validation regions~\cite{STDM-2017-30} and are fixed in the fit, except for $R^{\mathrm{iso(id)}}_{jj}$, which is allowed to float. The identification and isolation efficiencies are treated as free parameters for the leading (sub-leading) photon in the $j\gamma$ ($\gamma j$) component and for both photons in the $jj$ component, while all remaining parameters are fixed to their MC predictions. The probability functions for the electron and pile-up backgrounds, $f_{ee,i}$ and $f_{\mathrm{PU},i}$, are obtained from the dedicated estimates described below and are also fixed in the fit. Systematic uncertainties account for differences between the MC prediction and a data-driven estimate obtained in a validation region with relaxed isolation requirements.
The expected number of non-prompt events in region $i$ is:
\begin{equation}
\begin{aligned}
n_i^{\mathrm{non-prompt}}
&= N_{j\gamma} f_{j\gamma,i} +
N_{\gamma j}f_{\gamma j,i}  +N_{jj} f_{jj,i},
\end{aligned}
\end{equation}

where $N_{ \gamma j}$, $N_{j \gamma}$, and $N_{jj}$ are the total number of events from the non-prompt photon components over all the 16 regions and are extracted from the fit. The procedure is repeated independently for each bin of the EW-SR and QCD-CRs defined in Section~\ref{sec:ObjEvntSel}, and for each bin for the differential cross section measurements.

Another source of background are processes where electrons are misidentified as photons.   The main contribution to this background comes from $Z \rightarrow ee$, and in the EW-SR, the VBF production of $Z\rightarrow ee$ and $W\rightarrow e\nu$.  These background sources give a small contribution (0.6\% at most) and are estimated by using MC simulation. A data driven scale factor is applied to MC predictions to account for differences relative to data in the $e \rightarrow \gamma$ misidentification probability~\cite{PERF-2017-02}, and its full effect is propagated as a systematic uncertainty. The scale factor is applied only to reconstructed photon candidates that are matched to a particle-level electron. A global $5\%$ uncertainty is assigned to account for theoretical modelling uncertainties from the next-to-next-to-leading-order prediction of the $Z \rightarrow ee$ process. These uncertainties are found to have a negligible impact on the final results.

The last relevant background comes from \textit{pile-up photons}, where the photons are produced from different $pp$ collisions in the same bunch crossing. The pile-up background is obtained following the procedure used in Ref.~\cite{STDM-2017-30}, where the normalisation is estimated by fitting a two-component template to the distance of the $z-$origin of conversion tracks of the two photons ($z_{\gamma}$).  Only events where both photons are converted are considered, providing information about the direction.  The pile-up component is modelled as the convolution of the individual Gaussian distributions of $z_{\gamma}$. The measured pile-up fraction ranges from 1\% to 5\% depending on the region. Statistical uncertainties on the methods are near 100\% and therefore no additional systematic is added. The shape of the pile-up distribution is taken by overlaying on an event-by-event basis two different particle-level MC samples of inclusive single photon production, generated with \SHERPA[2.2.2] at NLO accuracy. For both this background and the electron background, these estimates are used to fix the corresponding probability density functions $f_{p,i}$ in the non-prompt fit.

Background events from Higgs boson production are modelled directly using MC simulation and constitute a minor contribution.


\section{Fiducial cross-section}
\label{sec:IncXS}

Two separate fits are performed, one to extract the EW-$\gamma\gamma jj$ cross-section ($\sigma_{\textrm{EW}} $) and one to extract the combined QCD+EW $\gamma\gamma jj$ cross-section ($\sigma_{\textrm{QCD+EW}} $).
In the EW-only case, a simultaneous binned maximum-likelihood fit to $m_{jj}$ in the EW-SR and QCD-CRs is performed to extract the signal strength $\mu_{\textrm{EW}} $.
The QCD-$\gamma\gamma jj$ background in the EW-SR is estimated with a two-dimensional sideband method~\cite{STDM-2010-08,D0:2010xdg,CDF:2011nuh}, to reduce the impact of possible mismodelling of \mjj. The total number of  QCD-$\gamma\gamma jj$ events in the EW-SR is obtained as:

\begin{equation}
N_{\mathrm{\gamma\gamma,SR}}^{\mathrm{QCD}}  = \mu^{\textrm{QCD}}_{\textrm{SR}}\cdot N^{\mathrm{MC,QCD}}_{\mathrm{\gamma\gamma,SR}},\\
\end{equation}

where $N^{\mathrm{MC,QCD}}_{\mathrm{\gamma\gamma,SR}} $ is the number of events as predicted from QCD-$\gamma\gamma jj$ MC simulations, and $ \mu^{\textrm{QCD}}_{\textrm{SR}} $ is the QCD yield scale factor in the EW-SR. The $ \mu^{\textrm{QCD}}_{\textrm{SR}} $ is obtained from the QCD-CR and is defined as:

\begin{equation}
\mu^{\textrm{QCD}}_{\textrm{SR}} = \frac{\mu^{\textrm{QCD}}_{\textrm{CR1}}\cdot \mu^{\textrm{QCD}}_{\textrm{CR2}}}{\mu^{\textrm{QCD}}_{\textrm{CR3}}},
\end{equation}
where, $\mu^{\textrm{QCD}}_{\textrm{CR1(2,3)}}$ denote the QCD-\yyjj yield scaling factors in the corresponding QCD-CR, obtained from a simultaneous fit together with $\mu_{\textrm{EW}}$.

In this procedure the fitted yield scaling factors constrain the QCD-\yyjj normalisation in the QCD-CR and the residual correlation between \xiyy and \Ngy is accounted through the MC prediction.
The significance of observing the EW-\yyjj process is evaluated using the profile likelihood ratio test statistic under the asymptotic approximation.
The EW-$\gamma\gamma jj$  cross-section is obtained by multiplying the signal strength $\mu^{\mathrm{EW}}$ by the cross-section predicted by the MC in the EW-SR region defined at particle-level.

The QCD+EW cross-section extraction is obtained with a similar method as in the EW-only, with a binned maximum-likelihood fit to $m_{jj}$,  combining the expected EW-\yyjj and QCD-\yyjj yields according to the MC predicted cross-sections.  The fit is performed only in the EW-SR and only one signal strength parameter ($\mu^{\textrm{QCD+EW}}$) is considered as parameter of interest in the fit. The fit is performed over $m_{jj}$, but with different bin boundaries than the EW-only fit. The QCD+EW cross-section is then obtained by multiplying the signal strength $\mu^{\textrm{QCD+EW}}$ by the expected cross-section in the EW-SR region defined at particle-level.


\section{Diﬀerential cross-sections}
\label{sec:DiffXS}
Events in the EW-SR are also used to measure differential cross-sections of EW-$\gamma\gamma jj$ and QCD+EW $\gamma\gamma jj$.
The detector-level differential distributions are corrected for detector effects such as limited resolution and imperfect selection efficiency, using a Profile Likelihood Unfolding technique, as implemented in the \texttt{TRExFitter} package~\cite{TRExFitter}.  The unfolding procedure uses the same likelihood function as that used in the inclusive measurement in Section~\ref{sec:IncXS}.
The number of particle-level events in a given bin $i$ ($N_i^{\mathrm{fid}}$) are related to the observed number of events ($N_j^{\mathrm{obs}}$) in a bin $j$ as
\begin{equation}
N_j^{\mathrm{obs}} = \frac{1}{A_j}\sum_i \epsilon_i M_{ij}N_i^{\mathrm{fid}}
\end{equation}

In this equation, $M_{ij}$ is the migration matrix, and represents the fraction of events at particle-level in bin $i$ that are reconstructed at detector-level in bin $j$. The $A_j$ is the acceptance correction, and it corrects for events that are in the EW-SR at detector-level, but do not enter the fiducial phase space at particle-level.
The $\epsilon_i$ term is the selection efficiency and it corrects for events entering the fiducial phase space at particle-level, that are not present in the detector-level EW-SR.
The migration matrix, acceptance, and efficiency corrections are obtained from MC simulations. In the EW-only differential measurements, only the EW-\yyjj MC is used, while for the QCD +EW differential measurements, both EW- and QCD-\yyjj MC are considered, combining the expected EW-\yyjj and QCD-\yyjj yields using the MC predicted cross-sections.

In the unfolding procedure,  each bin $i$ of the particle-level distribution is considered as a separate sample with its own signal-strength $\mu_{EW,i}$ (or $\mu_{QCD+EW,i}$). In the EW-only differential measurement, a simultaneous fit to the EW-SR and QCD-CRs allows to extract the $\mu_{EW,i}$ and the QCD-\yyjj contribution,  while in the QCD+EW differential measurements, only the EW-SR is considered in the fit and the $\mu_{QCD+EW,i}$ are extracted.
The differential cross-section measurements are performed for a set of observables, belonging in four groups. The first comprises photon kinematics and includes the transverse momenta of the two leading photons, together with the invariant mass and transverse momentum of the diphoton system.
The second category includes kinematic observables of the two leading jets: the transverse momenta of both jets and of their vectorial sum.
The third category targets the VBS topology and includes the dijet invariant mass, $m_{jj}$, the absolute rapidity separation of the two leading jets, $|\Delta y_{jj}|$, \Ngy, and the diphoton centrality, $\xi_{\gamma\gamma}$.
The final category contains observables sensitive to possible BSM effects: the signed azimuthal angle difference between the two leading jets, ordered in increasing $\eta$, $\Delta\phi_{jj}^{\mathrm{sign}}$, and its absolute value, $|\Delta\phi_{jj}|$. The observables $\xi_{\gamma\gamma}$ and \Ngy are unfolded only for the QCD+EW case. Since these observables enter the definition of the EW-SR, the fiducial phase space is extended only along the observable being unfolded. The binning of the observables is optimised to enhance the signal significance and purity and ensuring sufficiently low statistical uncertainty in each bin. The optimisation is done separately for the EW-only and the QCD+EW differential measurements.  When unfolding the $\mjj$ and $p_{\mathrm{T},j_{1(2)}}$  observables, an additional bin is included below the thresholds used to define the phase space, to correctly model the effects of migration at the low edge of the distribution. The results are however only shown inside the fiducial phase space.

%
%


\section{Systematic uncertainties}
\label{sec:uncertainties}
Different sources of systematic uncertainties affect the analysis in the expected yields or shapes. These are grouped in the following categories: experimental uncertainties,  theoretical uncertainties, and uncertainties in data-driven background modelling.

Experimental systematic uncertainties are related to the trigger, reconstruction, and identification of the final jets and photon candidates.  These uncertainties include also photon energy scale and resolution~\cite{PERF-2017-02}, jet energy scale and resolution~\cite{JETM-2018-05}, pile-up modelling~\cite{PERF-2014-03}, and the luminosity measurement~\cite{DAPR-2021-01} uncertainties.

In the EW-only fit, a systematic uncertainty is assigned to the MC predicted \xiyy and \Ngy correlation used in the QCD-\yyjj background estimate.
This systematic uncertainty is evaluated by comparing in simulation the nominal correlation with alternative two-dimensional sideband correlations. The first is based on QCD-CR2 and QCD-CR3 regions
($\Ngy=1$) together with dedicated validation regions ($\Ngy>1$ and $\mjj > 500~\GeV$) called QCD-VR1 ($\xiyy>0.5$ ) and QCD-VR2 ($\xiyy<0.5$). The other is based on the EW-SR and QCD-CR1–3 regions in the range of $250~\GeV < \mjj < 500~\GeV$. The first alternative differs from the nominal correction by 5\%, which is assigned as the systematic uncertainty, while the second provides a consistency cross-check yielding the same difference with respect to the nominal.

The systematic uncertainties related to the PDF choice set used in the sample generation are evaluated by considering 100 PDF variations and taking the standard deviation as the uncertainty. An additional uncertainty is obtained by varying up and down the strong coupling constant $\alpha_s$ in the nominal set and adding this in quadrature.
Uncertainties due to missing orders in the QCD calculations are evaluated by varying the renormalisation scale, $\mu_r$ and factorisation scale, $\mu_f$ by a factor of 0.5 or 2 individually,  avoiding however those cases where the difference between the scaling factors of $\mu_r$ and $\mu_f$ is larger than 2. The envelope of the six variations is taken as an uncertainty.
In both the EW-\yyjj and QCD+EW \yyjj cross-section fits, the scale uncertainty in the predicted \mjj distribution is decomposed into shape and overall normalisation components.
In the EW-\yyjj fit, the component associated with the normalisation of the EW-\yyjj sample is not kept while a nuisance parameter associated with the overall normalisation uncertainty for the QCD-\yyjj process is included and treated as correlated across the three QCD-CR regions. In the QCD+EW fit, the normalisation component of the uncertainty is dropped also for the QCD-\yyjj sample. Additional systematic uncertainties to account for different ME calculations and PS modelling are obtained by comparing the nominal EW-\yyjj and QCD-\yyjj samples with the respective alternative samples described in Section~\ref{sec:Samples}.
For this uncertainty, the same decomposition of shape and normalisation effects is performed as for the scale uncertainty.
In the EW-\yyjj fit, the interference between the EW and QCD production of \yyjj events is not included  in the fit. This contribution is computed with \MGNLO and compared with the contribution from the nominal EW-\yyjj sample. A difference of 5\% is observed, and a corresponding overall uncertainty is attributed to the signal.

To account for the statistical uncertainty of the MC samples an extra poisson uncertainty is added for each bin of each region. For the data-driven backgrounds, a separate nuisance parameter for each background category is considered accounting for the corresponding statistical uncertainty. The uncertainties related to the electrons, pile-up and $H\rightarrow\gamma\gamma$ backgrounds are as described in Section~\ref{sec:background}. The impact of the systematic uncertainties in the EW-\yyjj cross-section measurement (described in Section~\ref{sec:IncXS}) is shown in Table~\ref{tab:uncertainties}. The largest impact is coming from the EW-\yyjj QCD scale variations, followed by the statistical uncertainties in the estimation of the background.

\begin{table}[htbp]
\centering
\caption{Breakdown of the relative impact on the EW-\yyjj cross-section measurements of the systematic uncertainties, obtained from the covariance matrix. The uncertainties are grouped in different categories. The category 'Other backgrounds' includes the modelling of $ee$, pile-up and $H\rightarrow\gamma\gamma$ backgrounds.}
\renewcommand{\arraystretch}{1.2}
\begin{tabular}{l c}
\hline\hline
Source & Relative uncertainty [\%] \\
\hline
EW-\yyjj modelling & 13 \\
QCD-\yyjj modelling & 4.0 \\
Other backgrounds modelling  & 2.4  \\
Non-prompt stat.  uncertainty  & 8.7  \\
Jet & 5.0 \\
Photon & 2.5  \\
Pile-up modelling & 1.9 \\
Luminosity & 0.9  \\
MC statistical uncertainty & 3.8 \\
\hline
Total systematic uncertainty & 18 \\
Statistical uncertainty & 9.6 \\
\hline\hline
\end{tabular}
\label{tab:uncertainties}
\end{table}

For the differential cross-section measurement, all the systematic uncertainties described above are propagated through the unfolding procedure via the migration matrices. In addition, dedicated systematic uncertainties associated with the unfolding procedure are evaluated through a series of validation tests designed to assess its robustness. These uncertainties are derived by comparing the nominal unfolded results with those obtained using migration matrices constructed from alternative $\gamma\gamma jj$ MC predictions. A second set of tests evaluates the impact of the modelling by unfolding the data with migration matrices derived from MC templates reweighted to reproduce the distributions of the observed data. The difference in the result is included in the systematic uncertainty and is at most 5\%.


\section{Results}
\label{sec:results}
Figure~\ref{fig:Postfit} shows the results of the EW-only fit and of the QCD+EW fit. Table~\ref{tab:PostYield} shows the observed and predicted yields in the EW-SR and QCD-CRs, for EW-only and for the QCD+EW fit.  Good agreement between the post-fit predictions and the data is observed.  The three QCD yield scaling factors are:
\begin{equation}
\mu^{\textrm{QCD}}_{\textrm{CR1}} =0.64^{+0.40}_{-0.23},
\mu^{\textrm{QCD}}_{\textrm{CR2}} =0.77^{+0.49}_{-0.28},
\mu^{\textrm{QCD}}_{\textrm{CR3}} =0.52^{+0.35}_{-0.19}.
\end{equation}

The measured central values of the QCD yield scaling factors show an overall offset relative to the \SHERPA predictions; nevertheless the deviations from unity are not statistically significant. The largest contribution to the errors on the QCD yield scaling factors comes from scale uncertainties in the QCD-\yyjj prediction. The observed EW signal strength is $\mu^{\mathrm{EW}}_{\mathrm{obs}}=0.80^{+0.18}_{-0.15}$. This corresponds to an observed (expected) significance of 6.2 (7.8) $\sigma$  relative to the background-only hypothesis.  The observed fiducial EW-$\gamma\gamma jj$ cross-section is:

\begin{equation}
\sigma^{\mathrm{EW}}_{\mathrm{obs}}= 13.8^{+3.0}_{-2.6}\, \mathrm{fb}
\end{equation}

to be compared with the expected cross-section from \MGNLO of $\sigma = 17.1\pm 2.4\;\mathrm{fb}$.
The result of the EW+QCD fit gives a signal strength $\mu^{\mathrm{QCD+EW}}=0.83\pm 0.07$,   compatible within uncertainties with the combined EW-\yyjj and QCD-\yyjj yield scaling factors in the EW-only fit . This corresponds to an observed fiducial $\gamma\gamma jj$ cross-section of $\sigma_{\mathrm{obs}}^{\mathrm{QCD+EW}}=54.3^{+4.6}_{-4.8}\, \mathrm{fb}$. The expected fiducial cross-section from the combination of \MGNLO (EW-\yyjj) and \SHERPA  (QCD-\yyjj) is  $\sigma = 65.7\;^{+4.8}_{-5.0}\;\mathrm{fb}$.

\begin{figure}[htbp]
\centering
\subfloat[]{\includegraphics[width=0.35\textwidth]{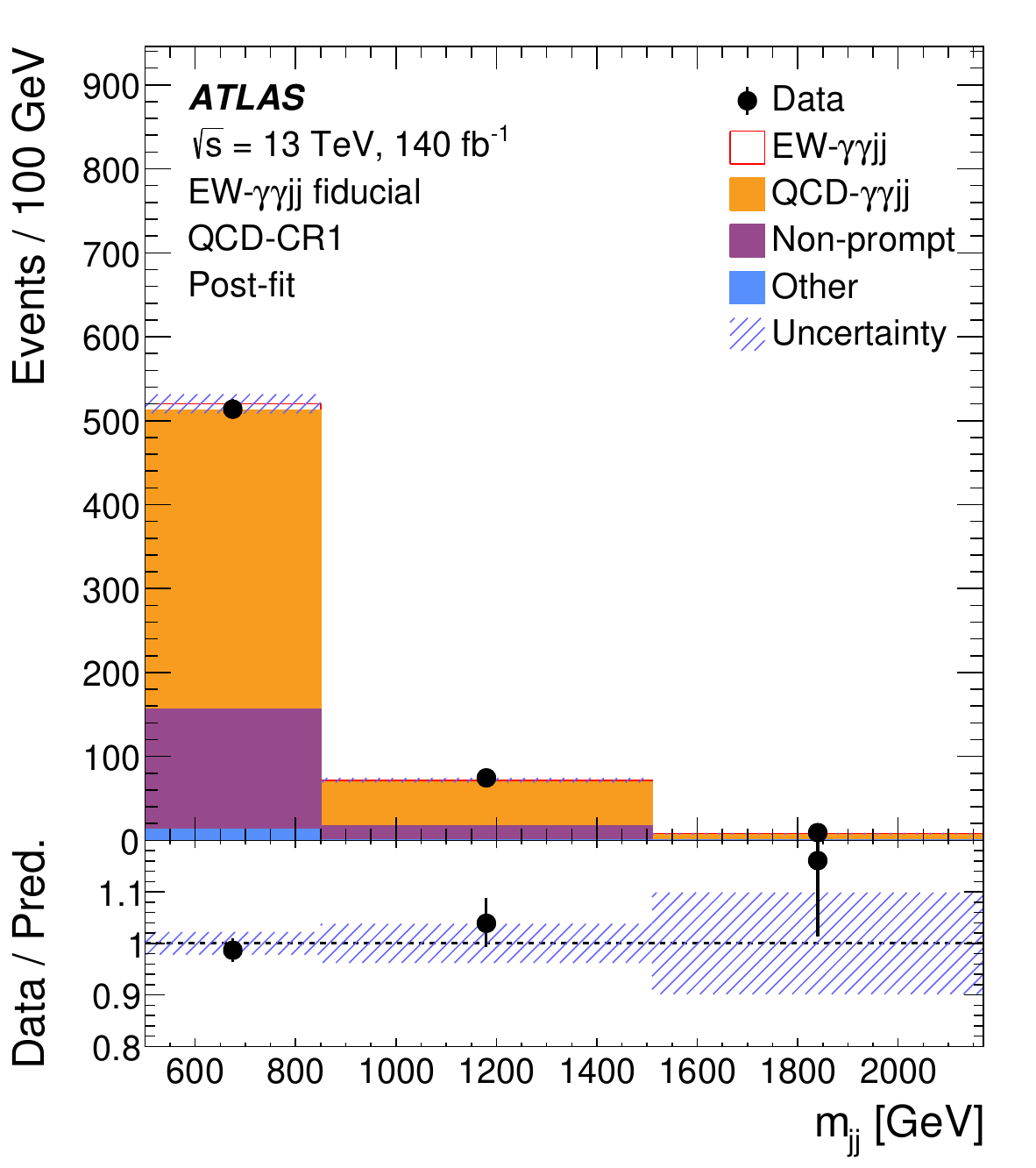}}
\subfloat[]{\includegraphics[width=0.35\textwidth]{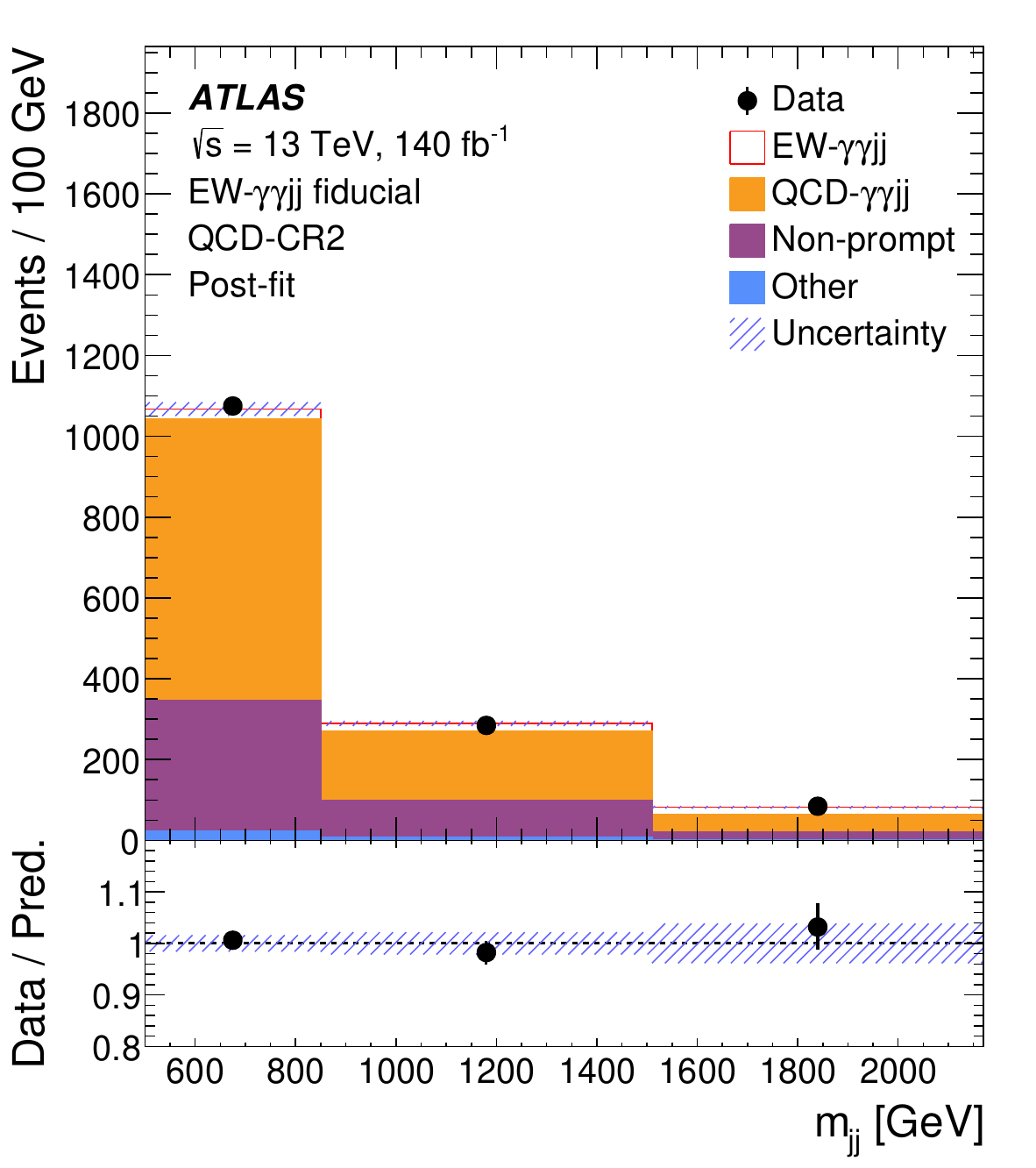}}
\subfloat[]{\includegraphics[width=0.35\textwidth]{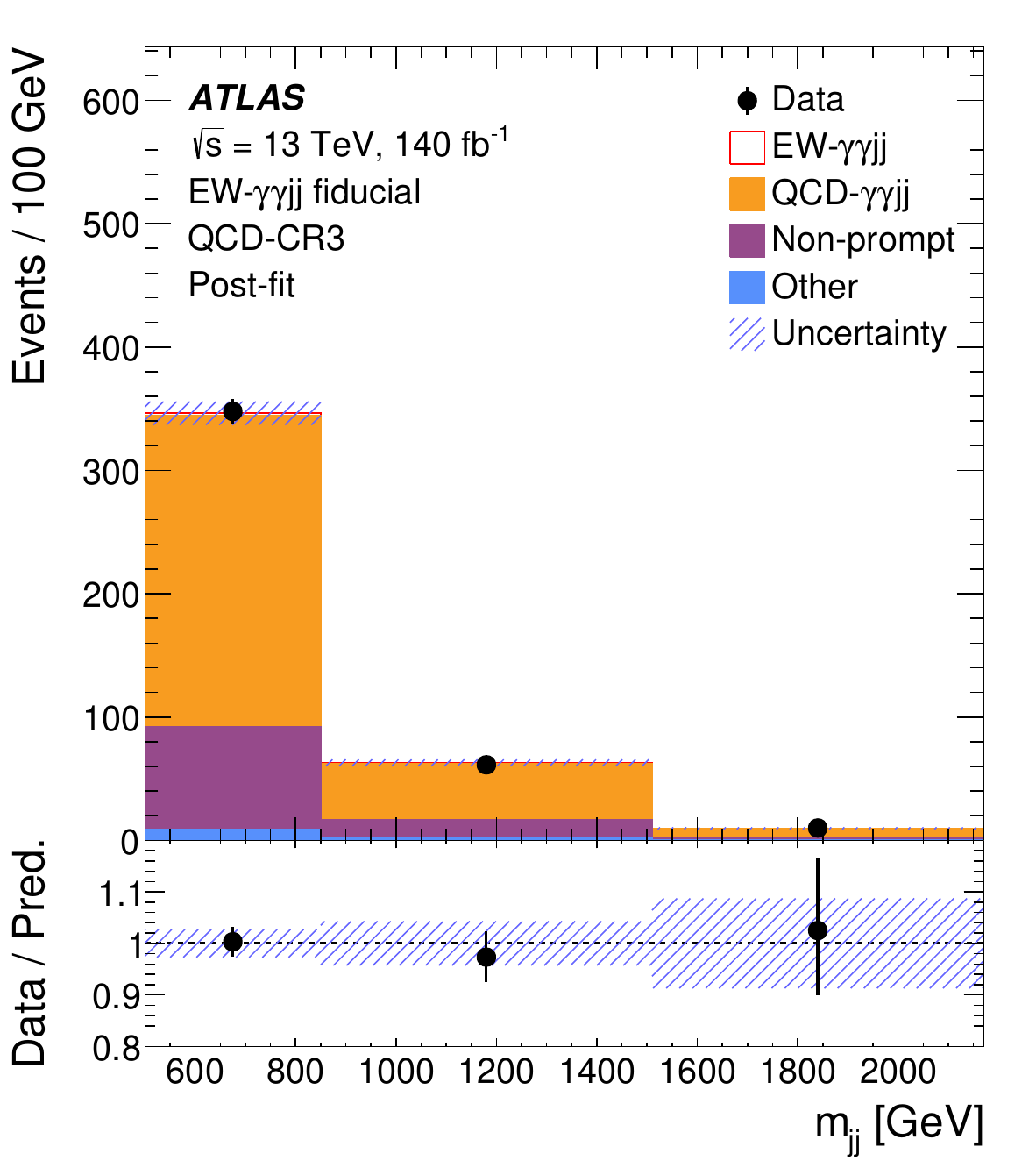}}\\
\subfloat[]{\includegraphics[width=0.35\textwidth]{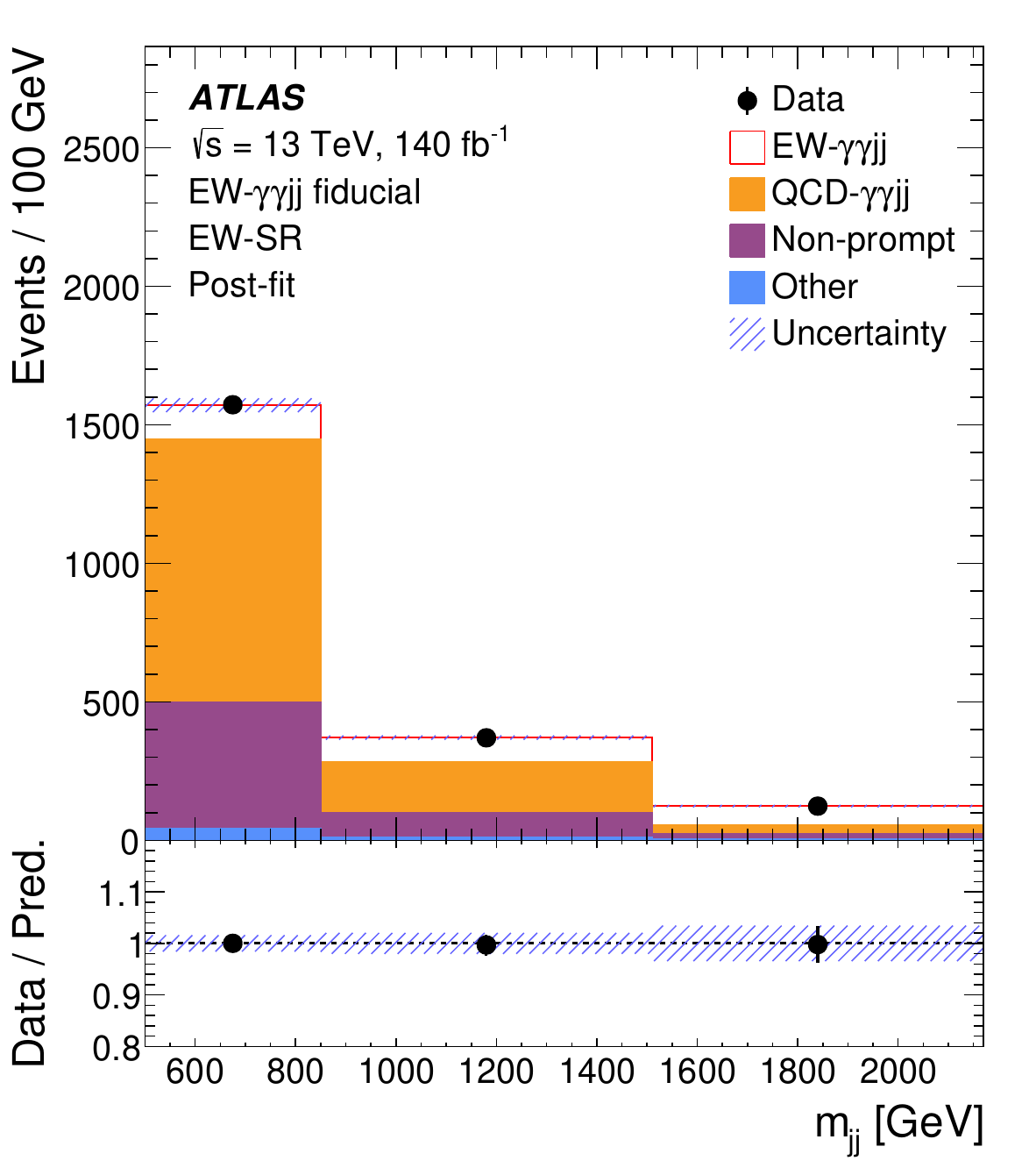}}
\subfloat[]{\includegraphics[width=0.35\textwidth]{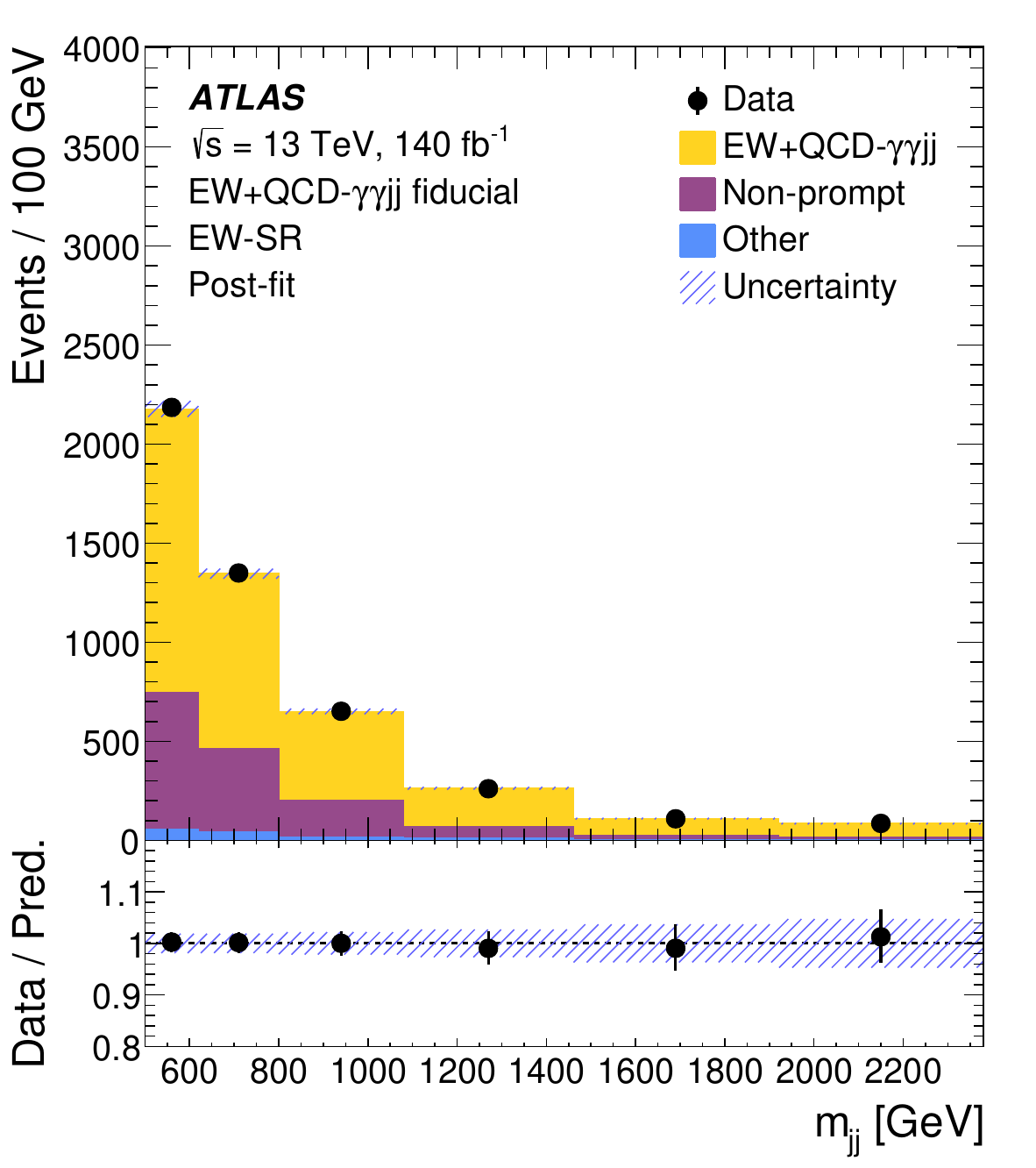}}\\
\caption{Post-fit distributions from the EW-only fit for $\mjj$ in (a) QCD-CR1, (b) QCD-CR2, (c) QCD-CR3, (d) EW-SR, and (e) of the QCD+EW fit in EW-SR. The dashed area represents the impact of all sources of systematic uncertainties and includes the correlations from the fit procedure. The bottom panel shows the ratio of the observed data to the MC predictions. Events beyond the upper limit of the histograms are included in the last bin.  }
\label{fig:Postfit}
\end{figure}

\begin{table}[!t]
\centering
\begin{small}
\caption{Summary table of the post-fit results for the EW-only fit in the EW-SR and QCD-CRs, and of the QCD+EW fit in the EW-SR.
Quoted uncertainties include all sources of systematic uncertainties and includes the correlations from the fit procedure.}
\label{tab:PostYield}

\begin{tabular}{
l
r @{${}\pm{}$} l
r @{${}\pm{}$} l
r @{${}\pm{}$} l
r @{${}\pm{}$} l
r @{${}\pm{}$} l
}
\toprule
& \multicolumn{8}{c}{EW-only fit}
& \multicolumn{2}{c}{EW+QCD fit}
\\
\midrule
& \multicolumn{2}{c}{QCD-CR1}
& \multicolumn{2}{c}{QCD-CR2}
& \multicolumn{2}{c}{QCD-CR3}
& \multicolumn{2}{c}{EW-SR}
& \multicolumn{2}{c}{EW-SR}
\\
\midrule

EW $\gamma\gamma jj$
& 46 & 12
& 330 & 70
& 10 & 5
& 1450 & 280
& \multicolumn{2}{c}{\multirow{2}{*}{6010$\pm$250}}
\\

QCD $\gamma\gamma jj$
& 1640 & 100
& 3850 & 220
& 1230 & 70
& 4740 & 340
& \multicolumn{2}{c}{}
\\

$\gamma j+j\gamma+jj$
& 610 & 90
& 1870 & 190
& 400 & 60
& 2310 & 220
& 2470 & 230
\\

electrons
& 18.3 & 2.3
& 25 & 7
& 13& 8
& 61 & 15
&61 & 15

\\

pile-up
& 30 & 10
& 100 & 30
& 36 & 6
& 150 & 26
& 150 & 26

\\
$H\rightarrow\gamma\gamma $
& 2.6 & 0.04
& 16.90 & 0.15
& 1.45 & 0.03
& 55.0 & 0.5
&55 & 0.5
\\

\hline

Total
& 2350 & 50
& 6190 & 80
& 1690 & 40
& 8760 & 110
& 8750 & 100
\\
\hline

Data
& \multicolumn{2}{c}{2348}
& \multicolumn{2}{c}{6192}
& \multicolumn{2}{c}{1686}
& \multicolumn{2}{c}{8754}
& \multicolumn{2}{c}{8754}
\\

\bottomrule
\end{tabular}
\end{small}
\end{table}

The measured differential cross-sections for the combined QCD+EW $\gamma\gamma jj$ process are presented in Figures~\ref{fig:EQResult1} and~\ref{fig:EQResult2}. The unfolded data are compared with the prediction from  \SHERPA for the QCD contribution and \MGNLO interfaced with \PYTHIA for the EW contribution, and with alternative configurations in which both the QCD and EW contributions are modelled entirely with  \MGNLO. Overall, both generator configurations provide a good description of the measured distributions within the experimental and theoretical uncertainties. The invariant-mass and transverse-momentum distributions of both the dijet and diphoton systems are generally well described by both \SHERPA and \MGNLO within uncertainties.
The two observables used to define the EW-SR and QCD-CR, $\xi_{\gamma\gamma}$ and \Ngy, are well described within the EW-SR by both prediction configurations.  Outside the EW-SR, the \SHERPA QCD-$\gamma\gamma jj$ predictions exhibit more noticeable differences with respect to the data in the \Ngy and \xiyy distributions.
The total uncertainties on the measured differential cross-sections vary from approximately 10\% to 40\%, depending on the observable and bin.

\begin{figure}[!htbp]
\centering
\subfloat[]{\includegraphics[width=0.33\textwidth]{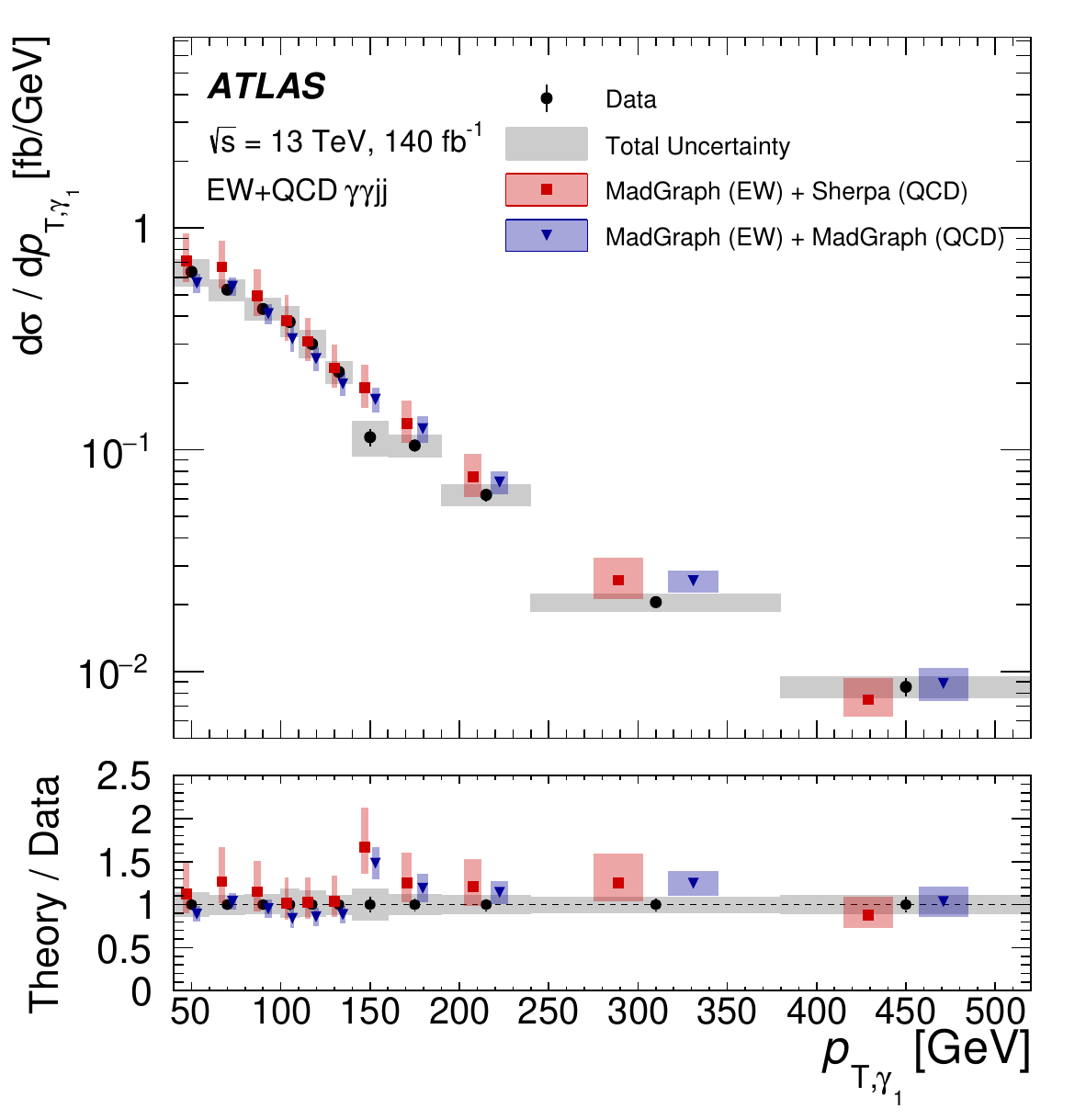}}
\subfloat[]{\includegraphics[width=0.33\textwidth]{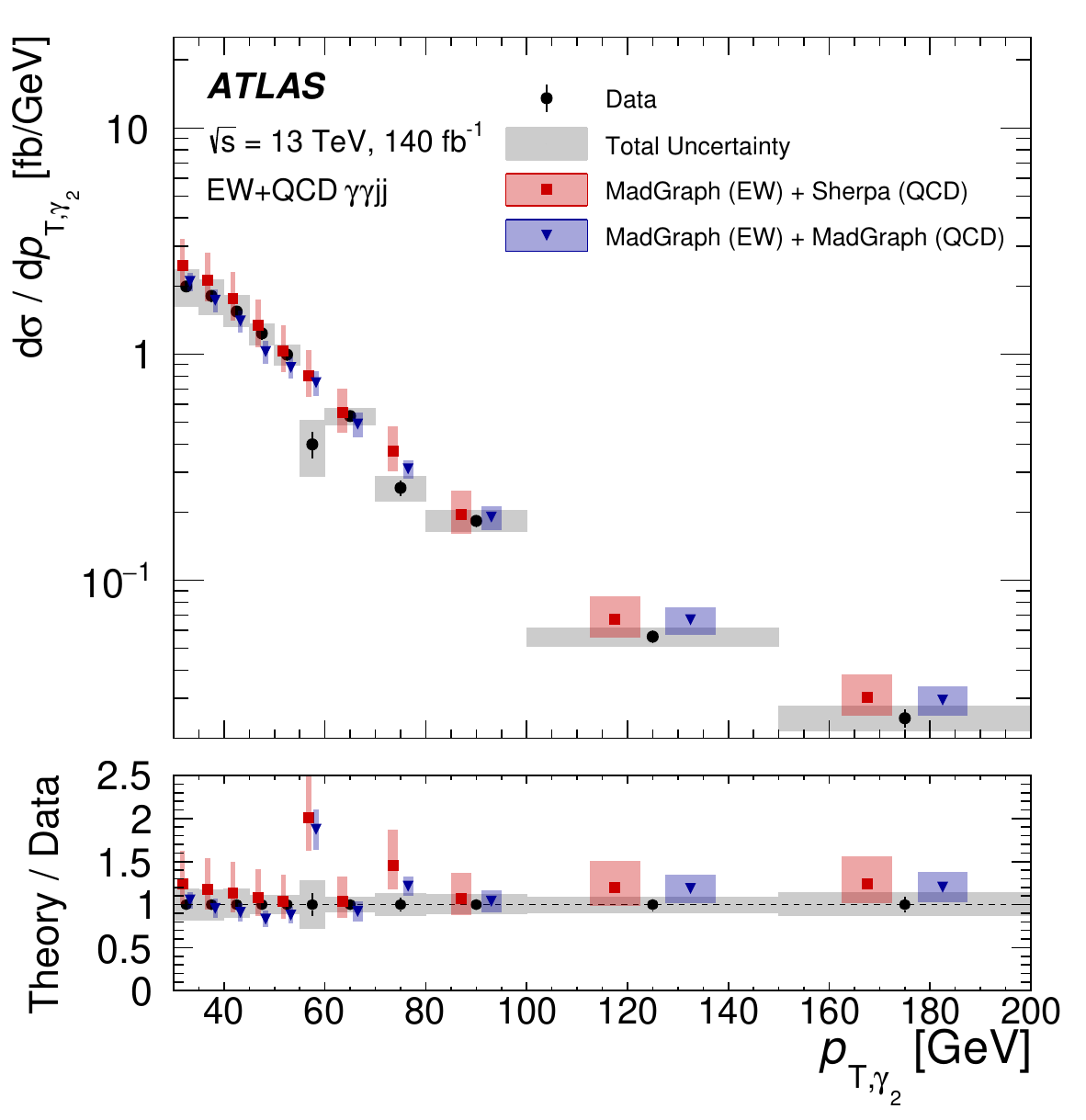}} \\
\subfloat[]{\includegraphics[width=0.33\textwidth]{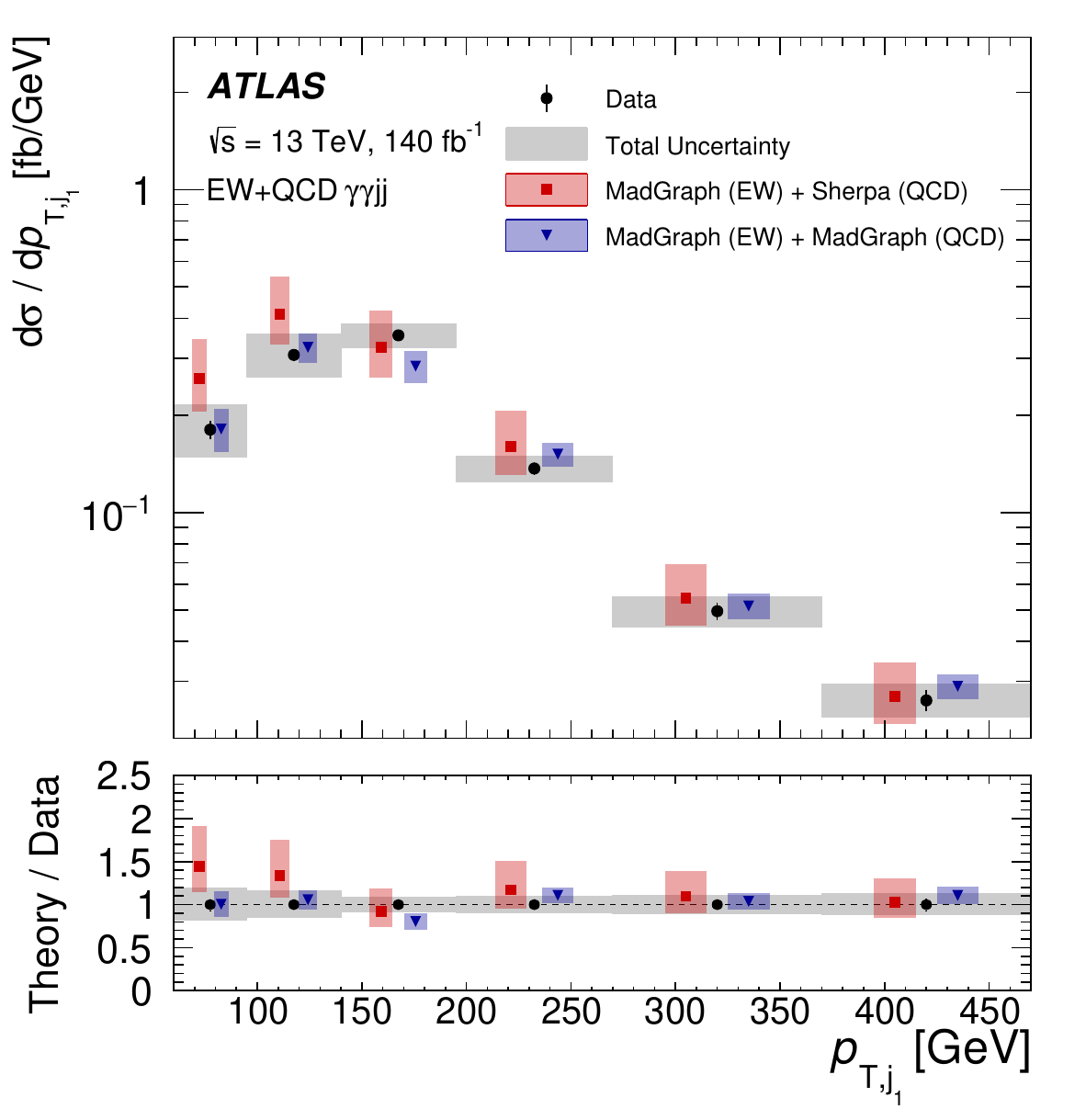}}
\subfloat[]{\includegraphics[width=0.33\textwidth]{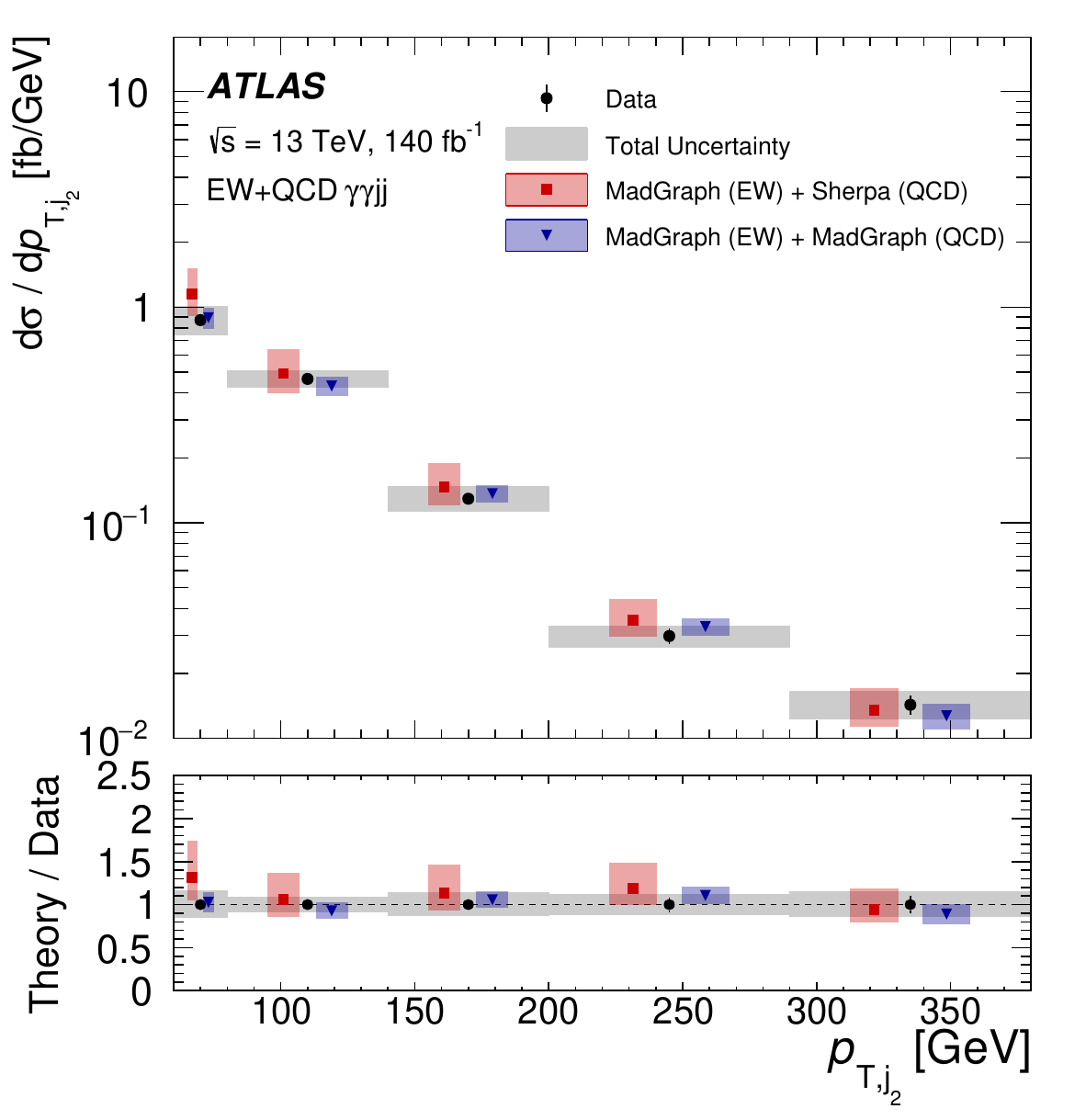}}\\
\caption{Differential QCD+EW $\gamma\gamma jj$ measurements for:
(a) $p_{\mathrm{T},\gamma_1}$,
(b) $p_{\mathrm{T},\gamma_2}$,
(c) $p_{\mathrm{T},j_1}$, and
(d) $p_{\mathrm{T},j_2}$.
The lower panel shows the ratio of the predictions to the data.
The black dots represent the data points with the statistical uncertainties, while the grey band represents the total uncertainties.
Particle-level predictions of $\gamma\gamma jj$ are shown for \MGNLO  interfaced with \PYTHIA (EW contribution) and \SHERPA  (QCD contribution) and for \MGNLO  interfaced with \PYTHIA for both EW and QCD contribution.  The coloured area around the predictions represents the effect of the systematic uncertainties from the renormalisation and factorisation scales. The last bin includes overflow events.}
\label{fig:EQResult1}
\end{figure}

\begin{figure}[!htbp]
\centering
\subfloat[]{\includegraphics[width=0.33\textwidth]{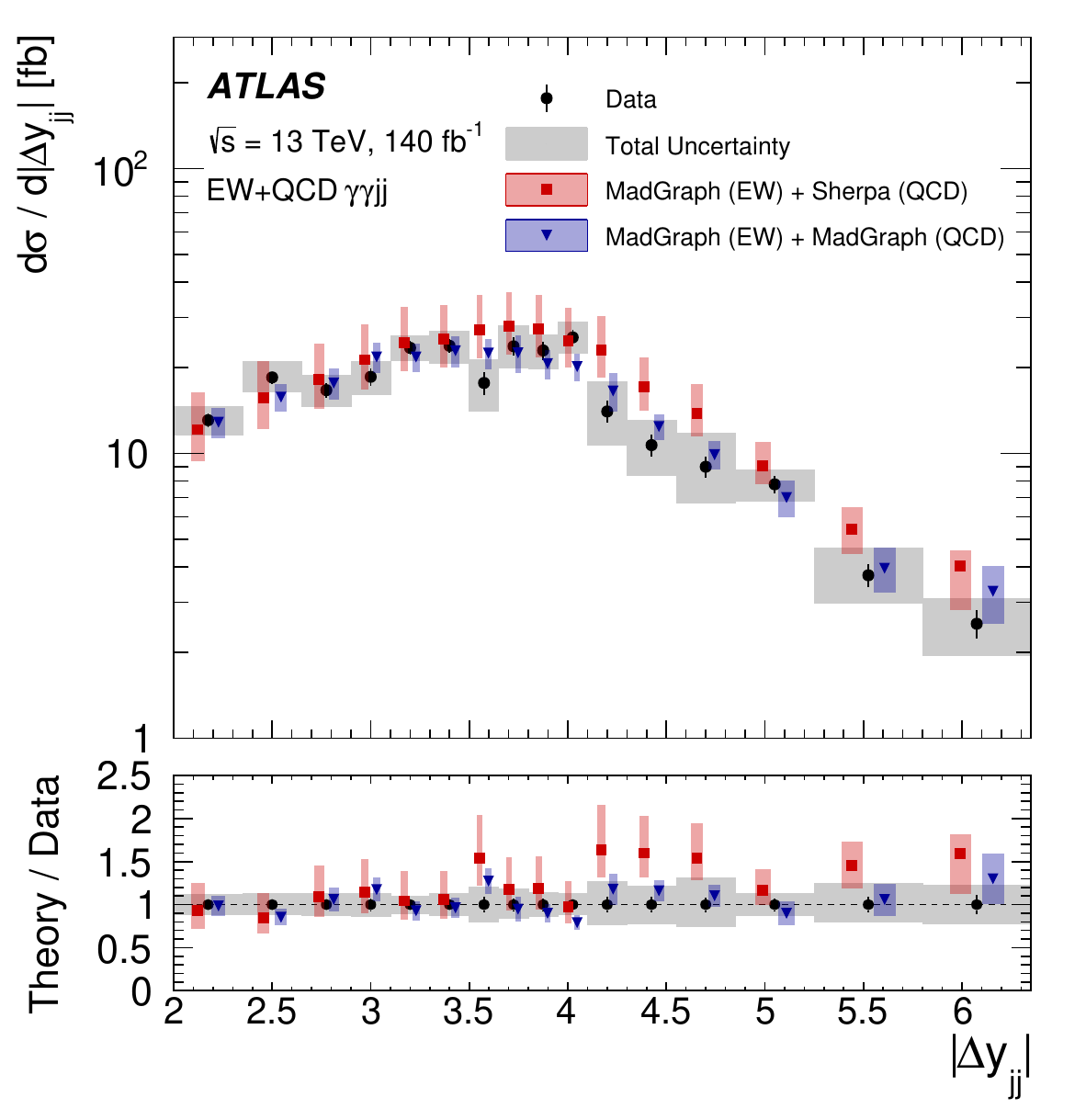}}
\subfloat[]{\includegraphics[width=0.33\textwidth]{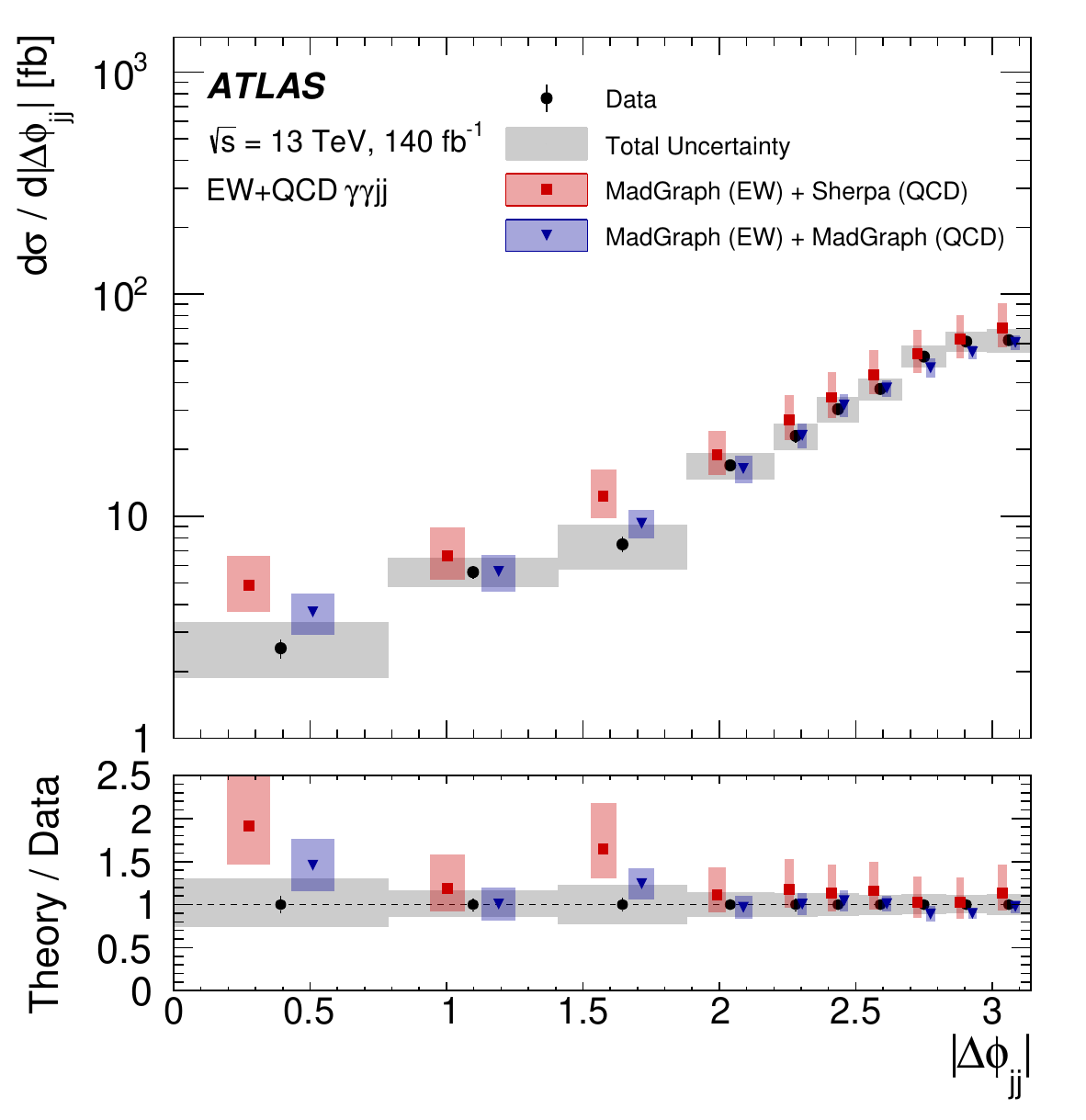}}
\subfloat[]{\includegraphics[width=0.33\textwidth]{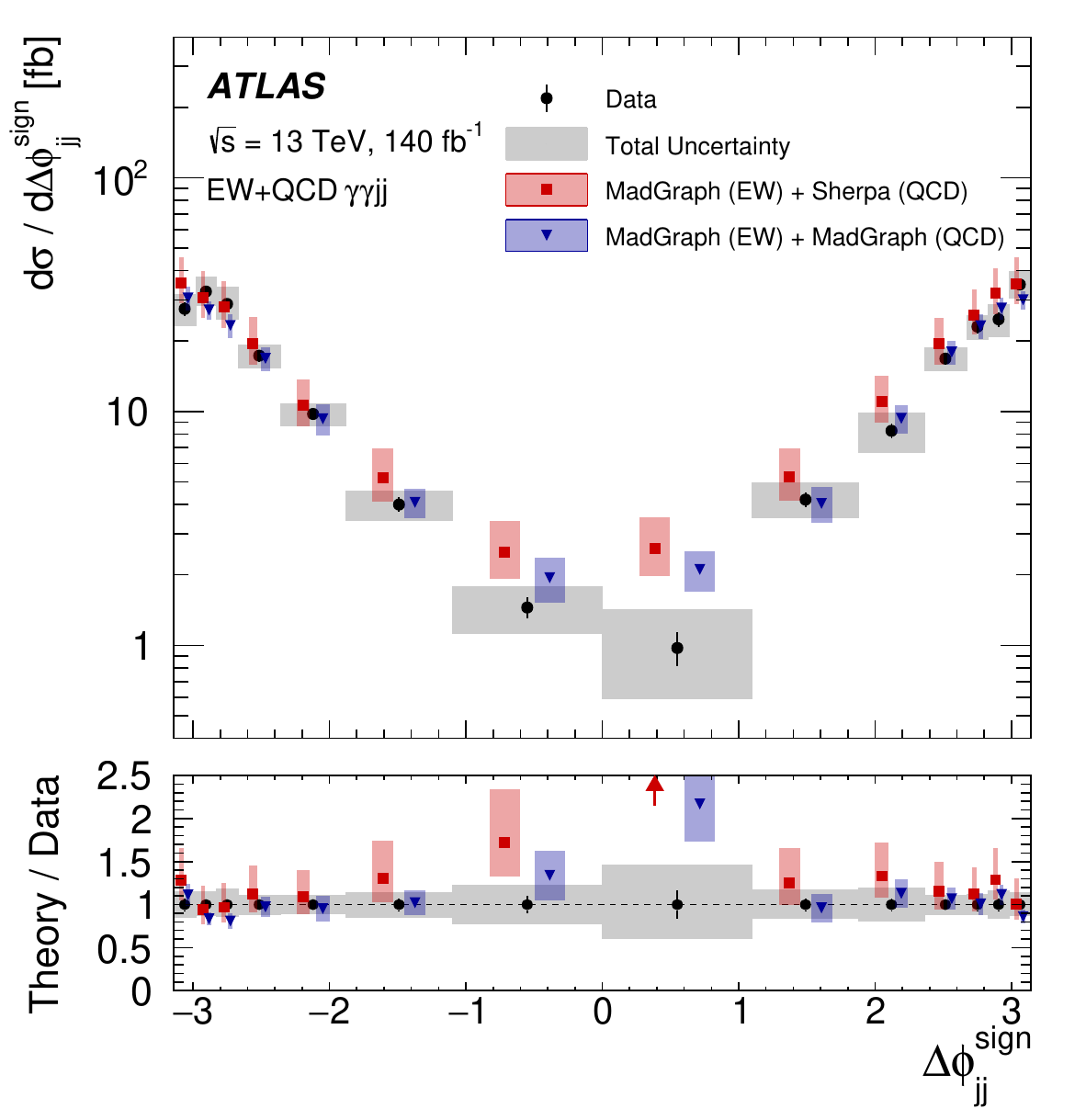}}\\

\subfloat[]{\includegraphics[width=0.33\textwidth]{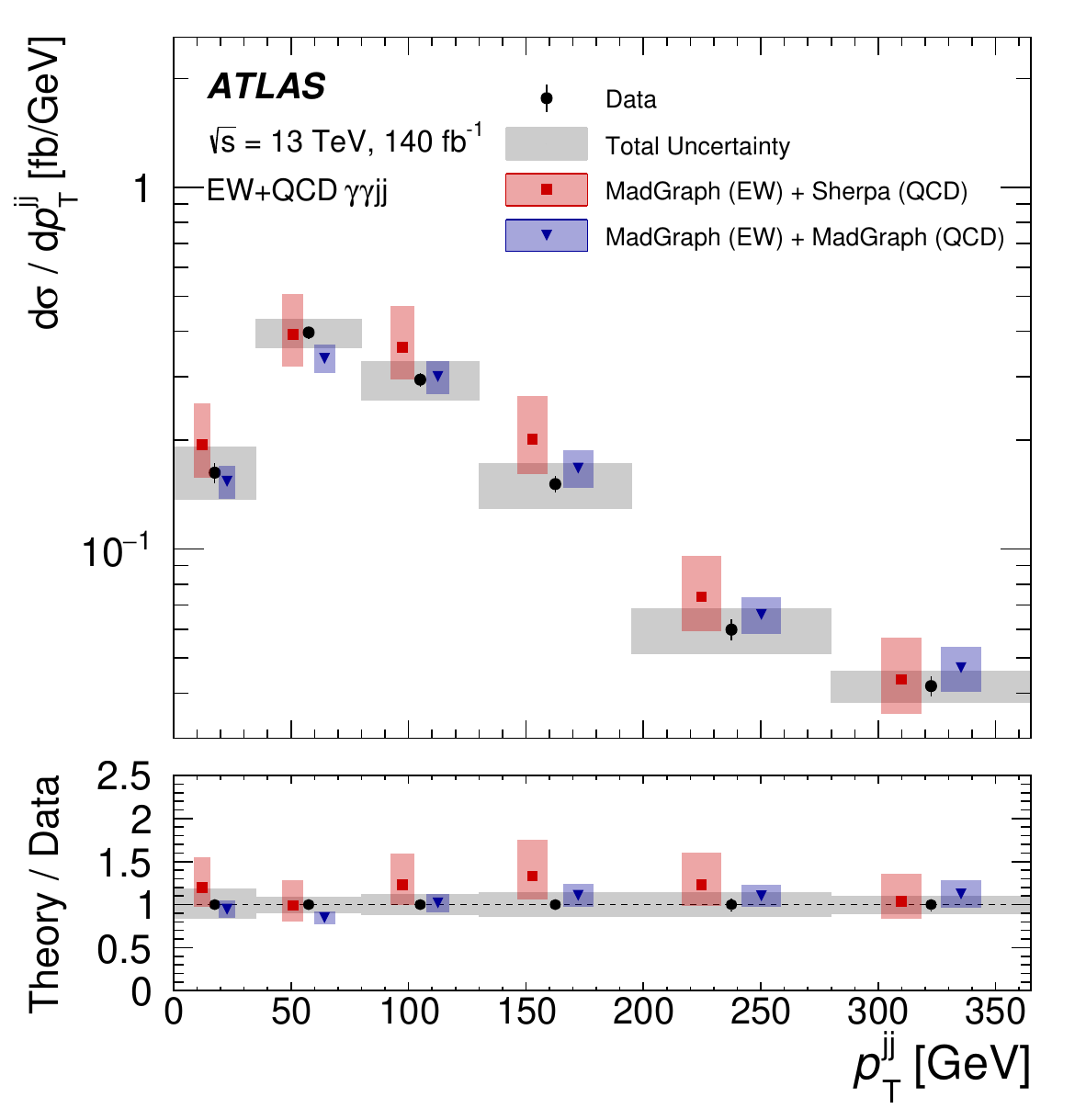}}
\subfloat[]{\includegraphics[width=0.33\textwidth]{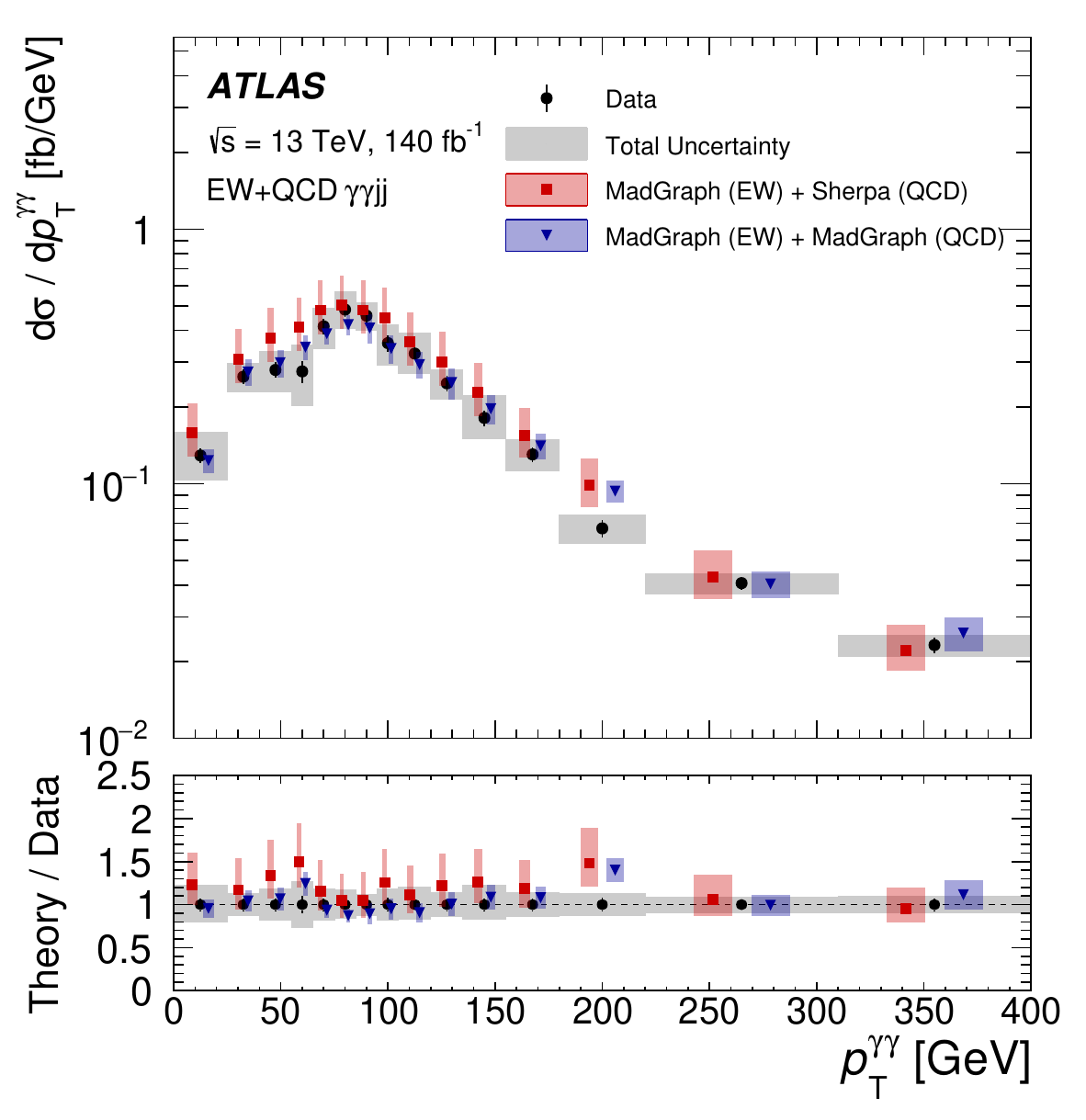}}
\subfloat[]{\includegraphics[width=0.33\textwidth]{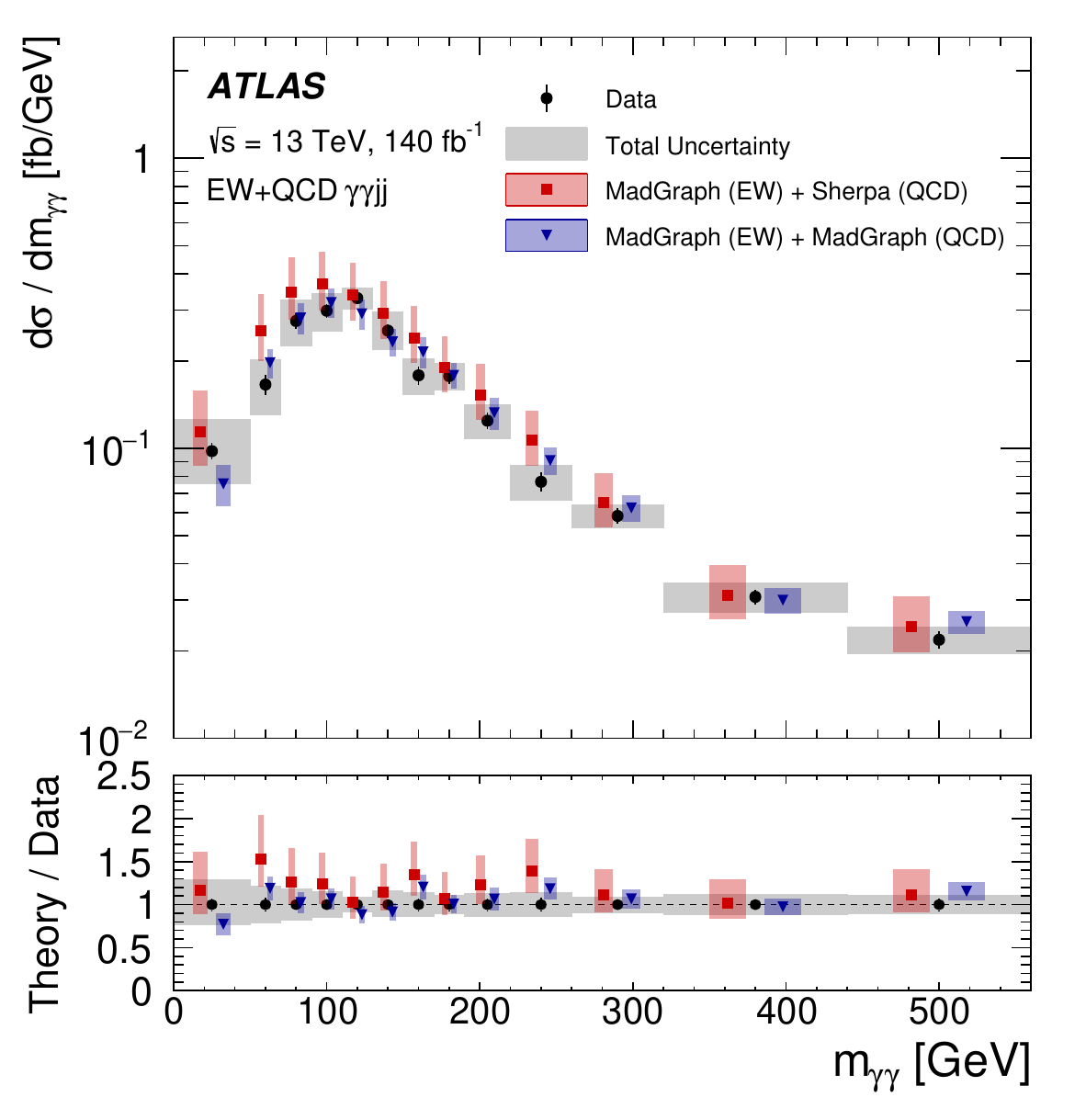}} \\

\subfloat[]{\includegraphics[width=0.33\textwidth]{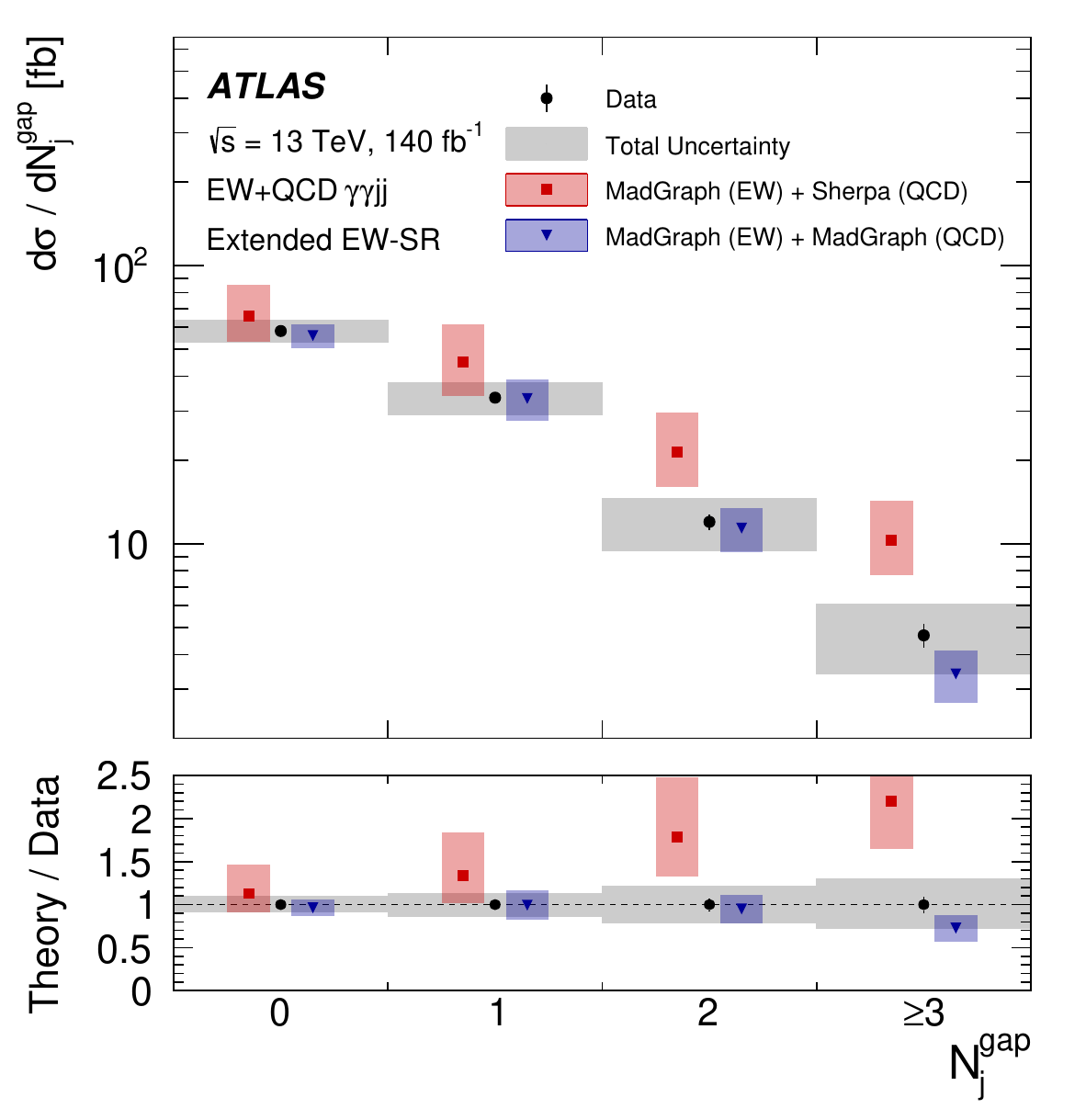}}
\subfloat[]{\includegraphics[width=0.33\textwidth]{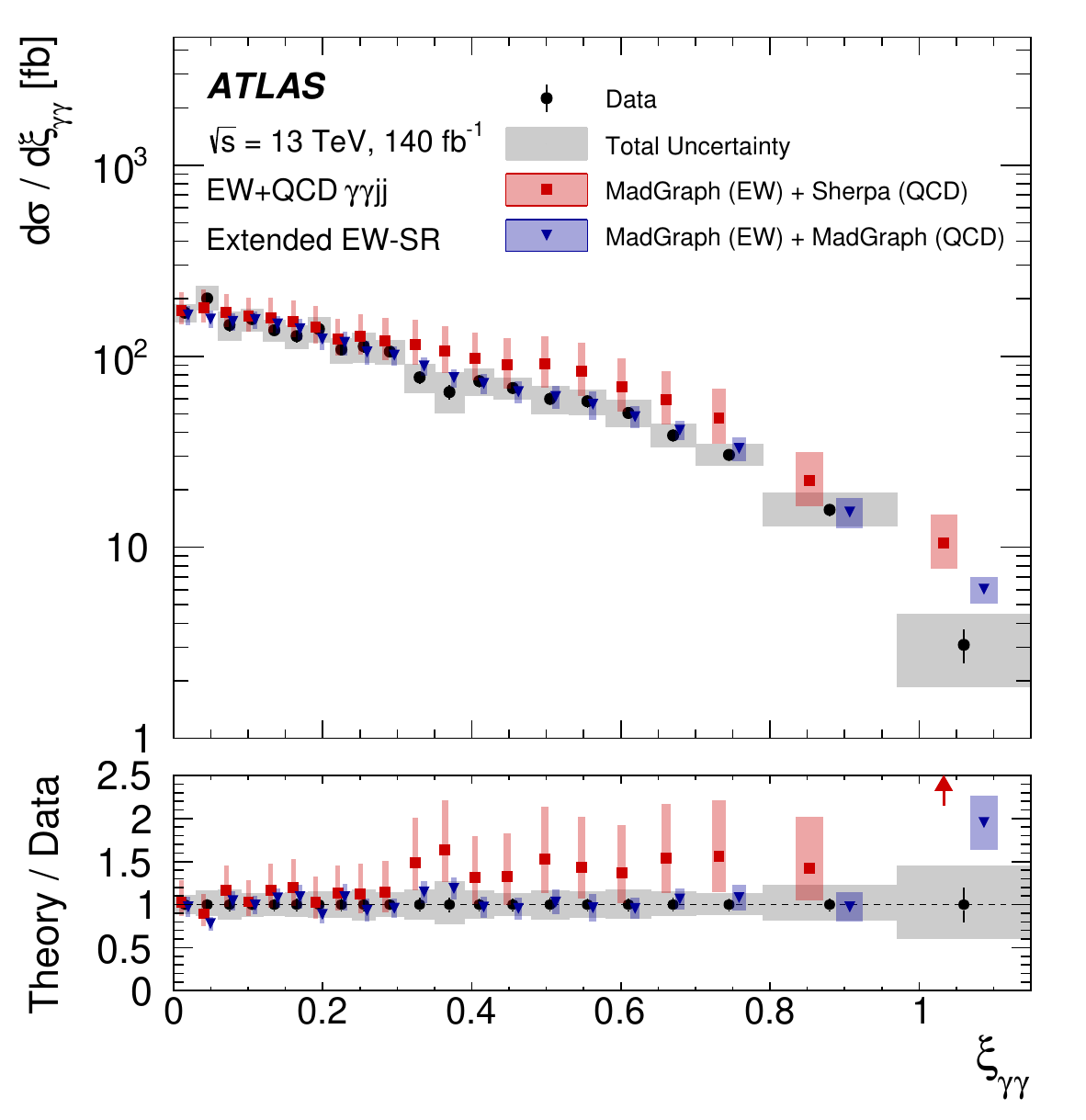}}
\subfloat[]{\includegraphics[width=0.33\textwidth]{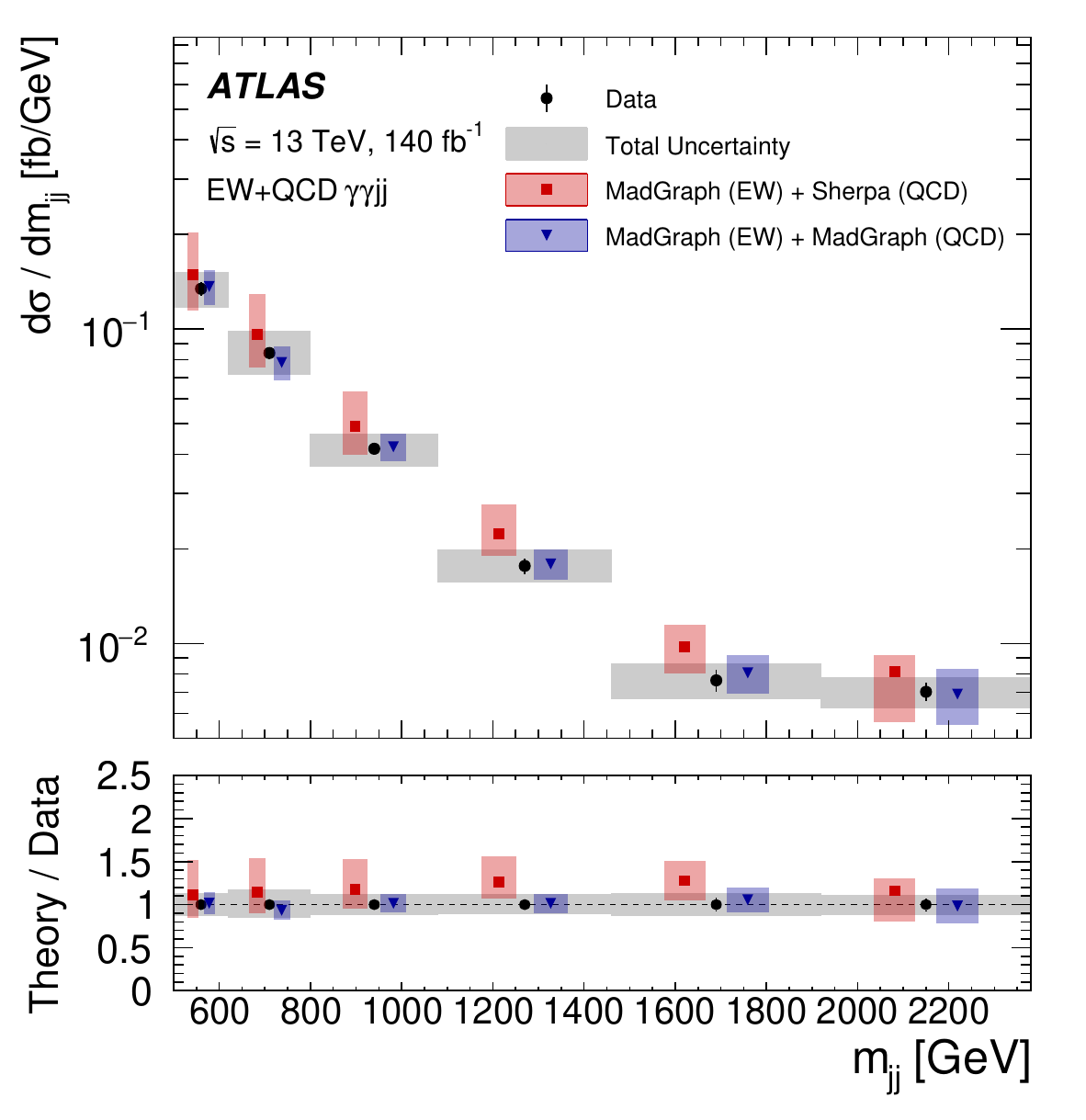}} \\

\caption{Differential QCD+EW $\gamma\gamma jj$ measurements for:
(a) $\Delta y_{jj}$,
(b) $|\Delta\phi_{jj}|$,
(d) $\Delta\phi_{jj}^{\text{sign}}$,
(c) $p_{\mathrm{T},jj}$,
(e) $p_{\mathrm{T},\gamma\gamma}$,
(f) $\myy$,
(g) $\Ngy$,
(h) $\xi_{\gamma\gamma}$ and
(i)  $\mjj$.
The lower panel shows the ratio of the predictions to the data.
The black dots represent the data points with the statistical uncertainties, while the grey band represents the total uncertainties.
Particle-level predictions of $\gamma\gamma jj$ are shown for \MGNLO  interfaced with \PYTHIA (EW contribution) and \SHERPA  (QCD contribution) and for \MGNLO  interfaced with \PYTHIA for both EW and QCD contribution. The coloured area around the predictions represents the effect of the systematic uncertainties from the renormalisation and factorisation scales. Except for $|\Delta\phi_{jj}|$ and $\Delta\phi_{jj}^{\text{sign}}$ the last bin includes overflow events.}
\label{fig:EQResult2}
\end{figure}

The measured differential cross-sections for the EW-$\gamma\gamma jj$ process are presented in Figure~\ref{fig:EResults1}, representing the first such measurement at the LHC. Overall, the unfolded data are in good agreement with the predictions. The differential cross-sections measurements are compared with the EW-$\gamma\gamma jj$ predictions from both \SHERPA and \MGNLO. Both predictions show similar behaviours relative to the data.
The uncertainties range from approximately 20\% (in the highest-\mjj bin) to over 100\%, and are mainly driven by the non-prompt photon statistical uncertainty, together with modelling uncertainties in the QCD-$\gamma\gamma jj$ background. The smallest uncertainties are achieved for \mjj and $p_{\mathrm{T},j_2}$.

\begin{figure}[!htbp]
\centering
\subfloat[]{\includegraphics[width=0.33\textwidth]{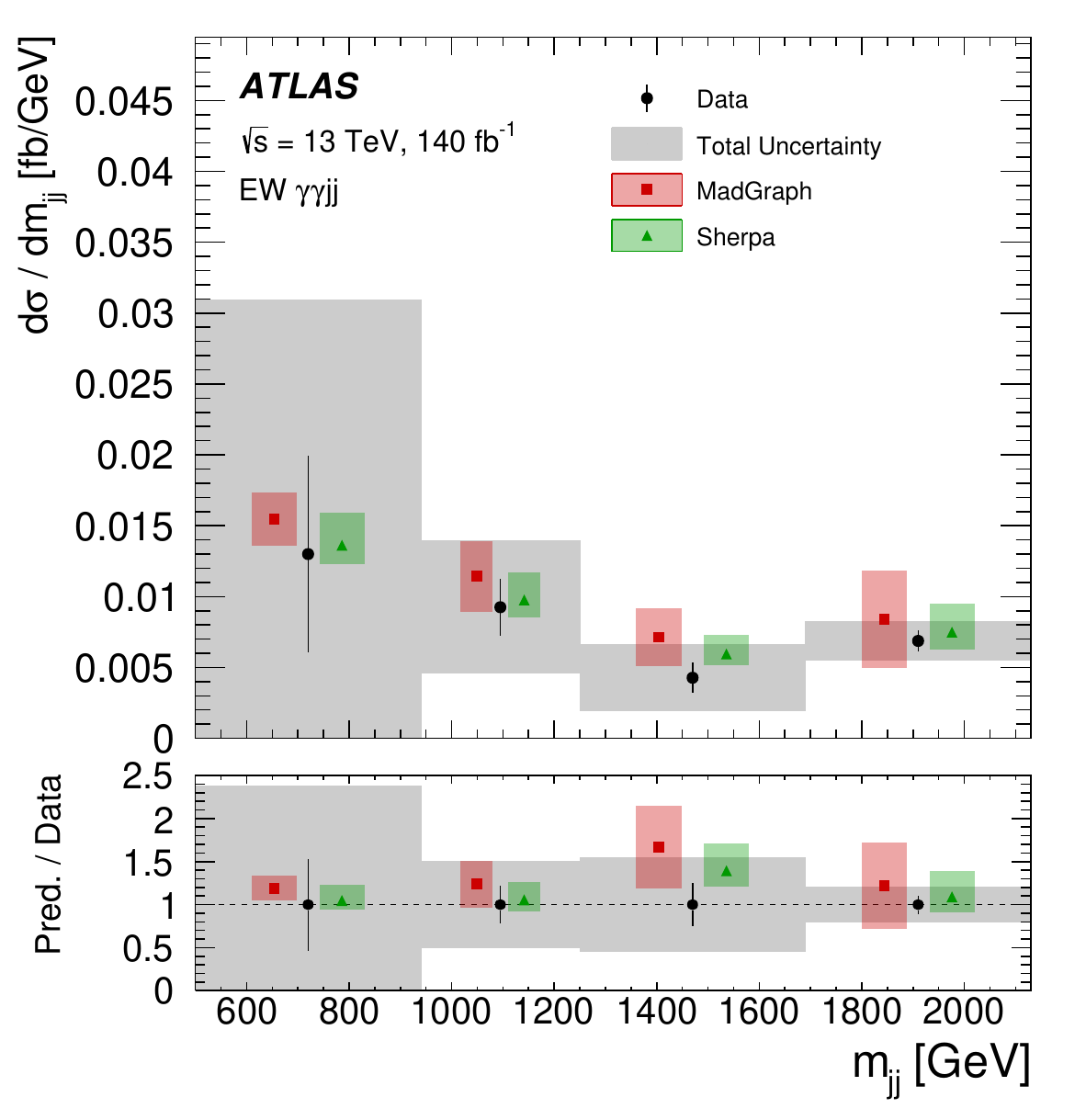}}
\subfloat[]{\includegraphics[width=0.33\textwidth]{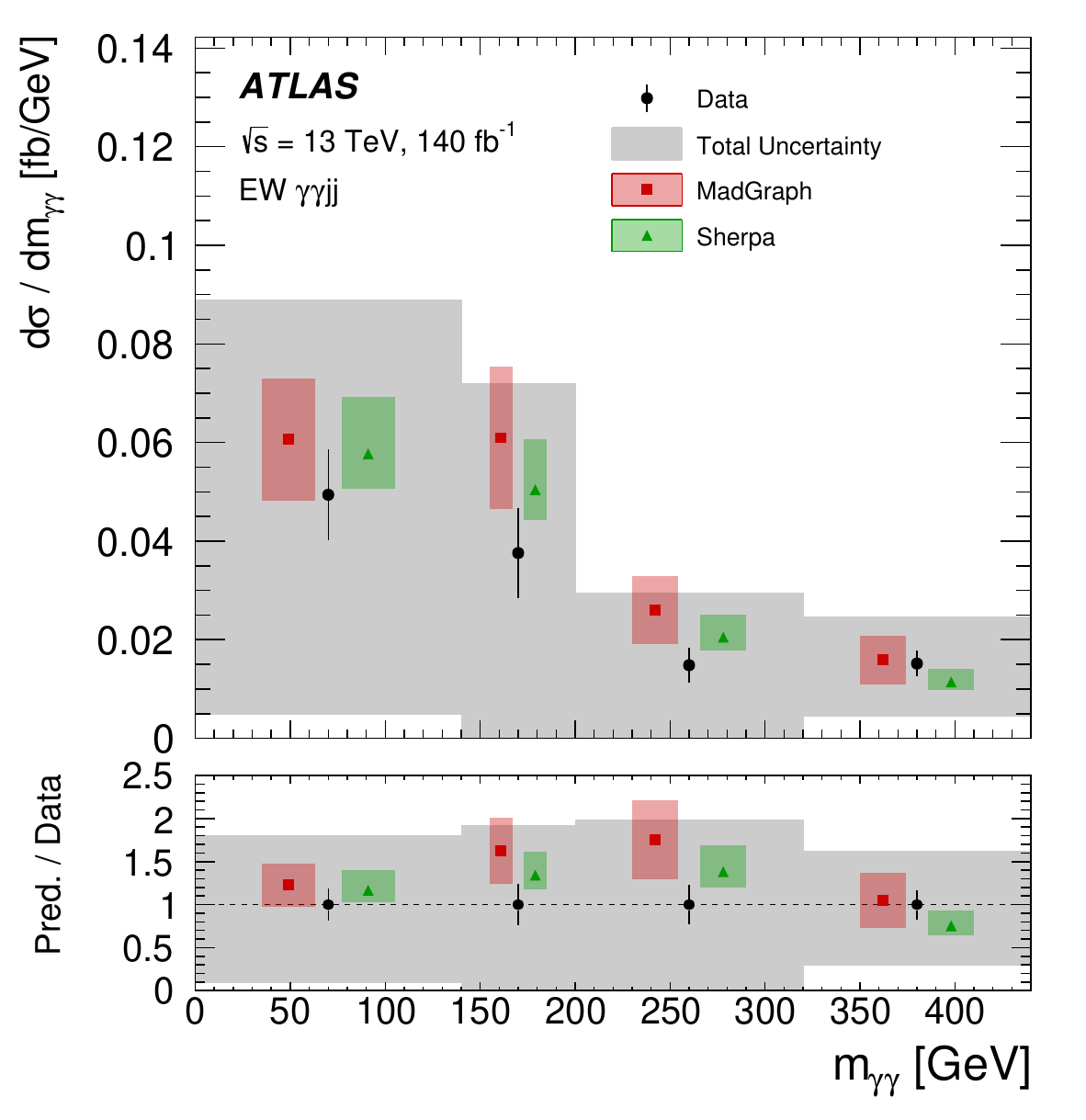}}
\subfloat[]{\includegraphics[width=0.33\textwidth]{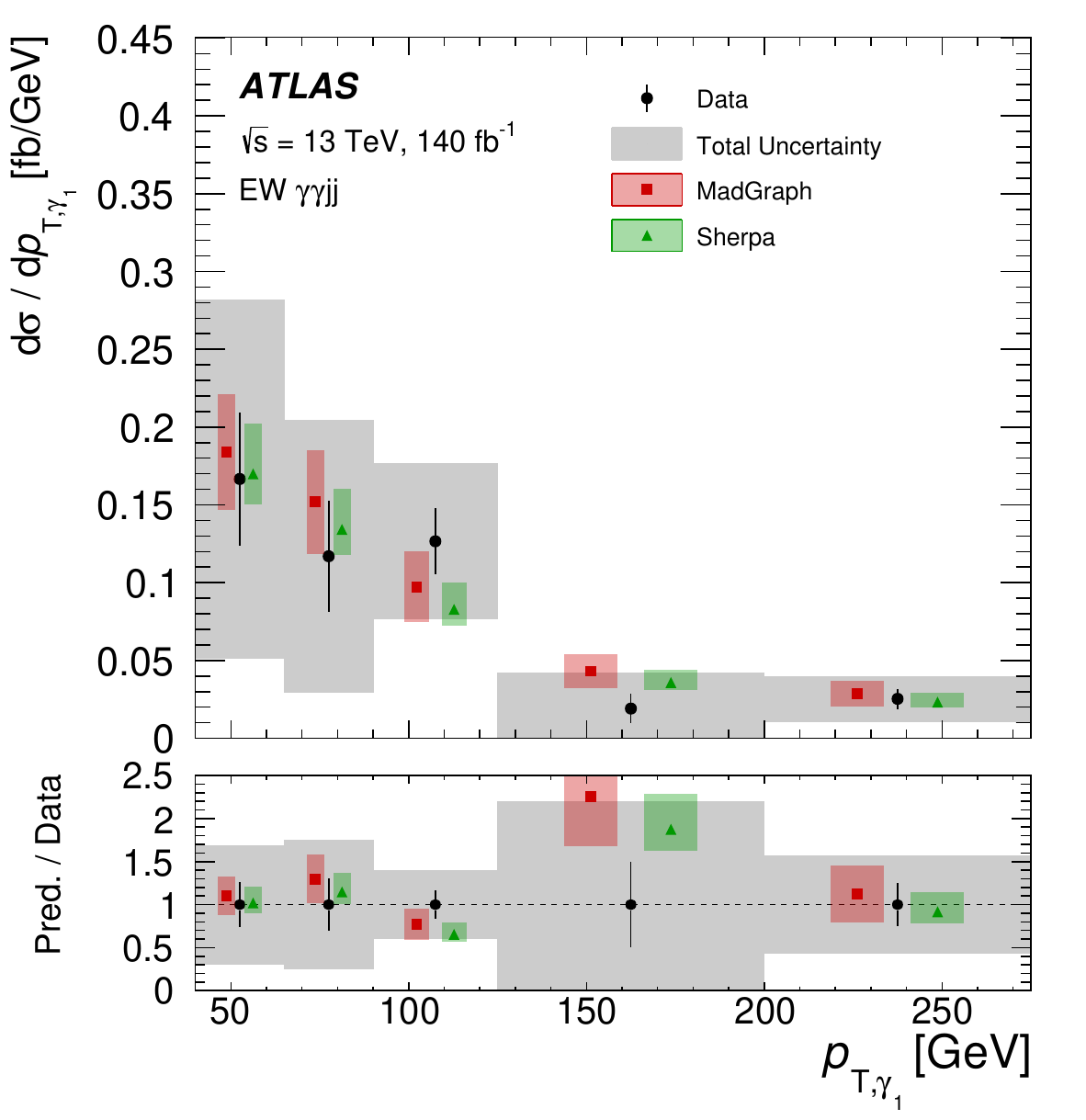}}\\
\subfloat[]{\includegraphics[width=0.33\textwidth]{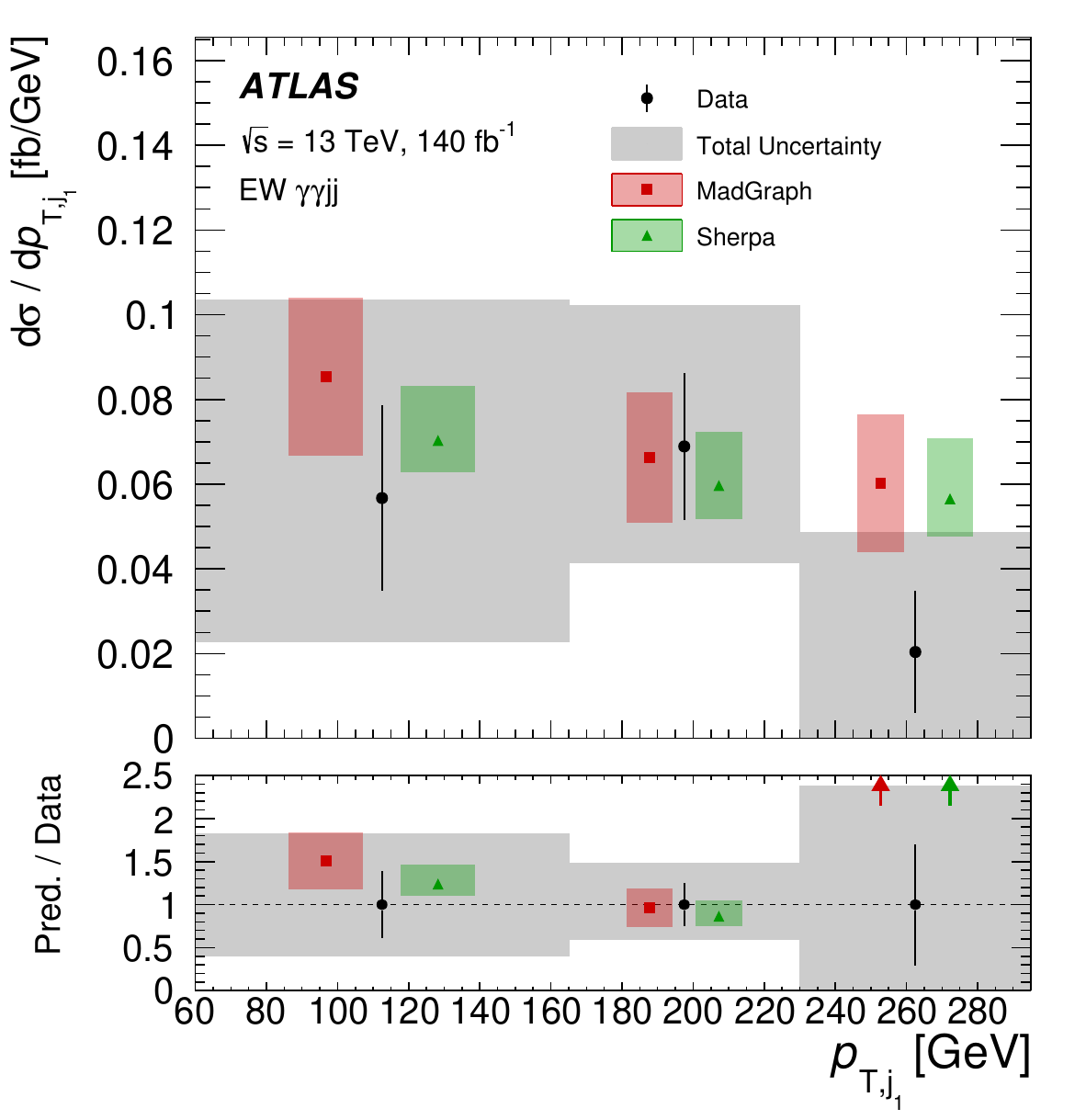}}
\subfloat[]{\includegraphics[width=0.33\textwidth]{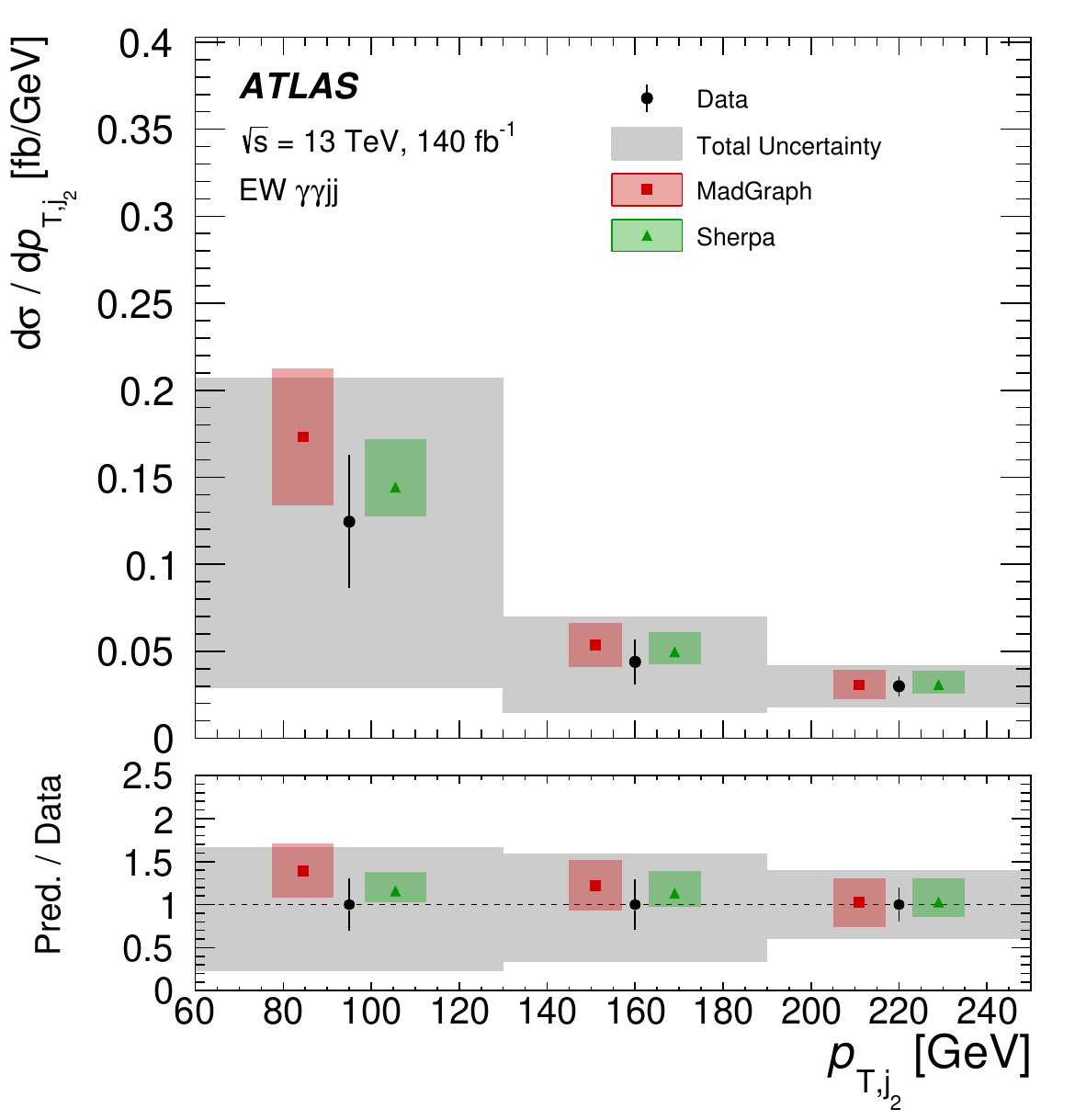}}
\subfloat[]{\includegraphics[width=0.33\textwidth]{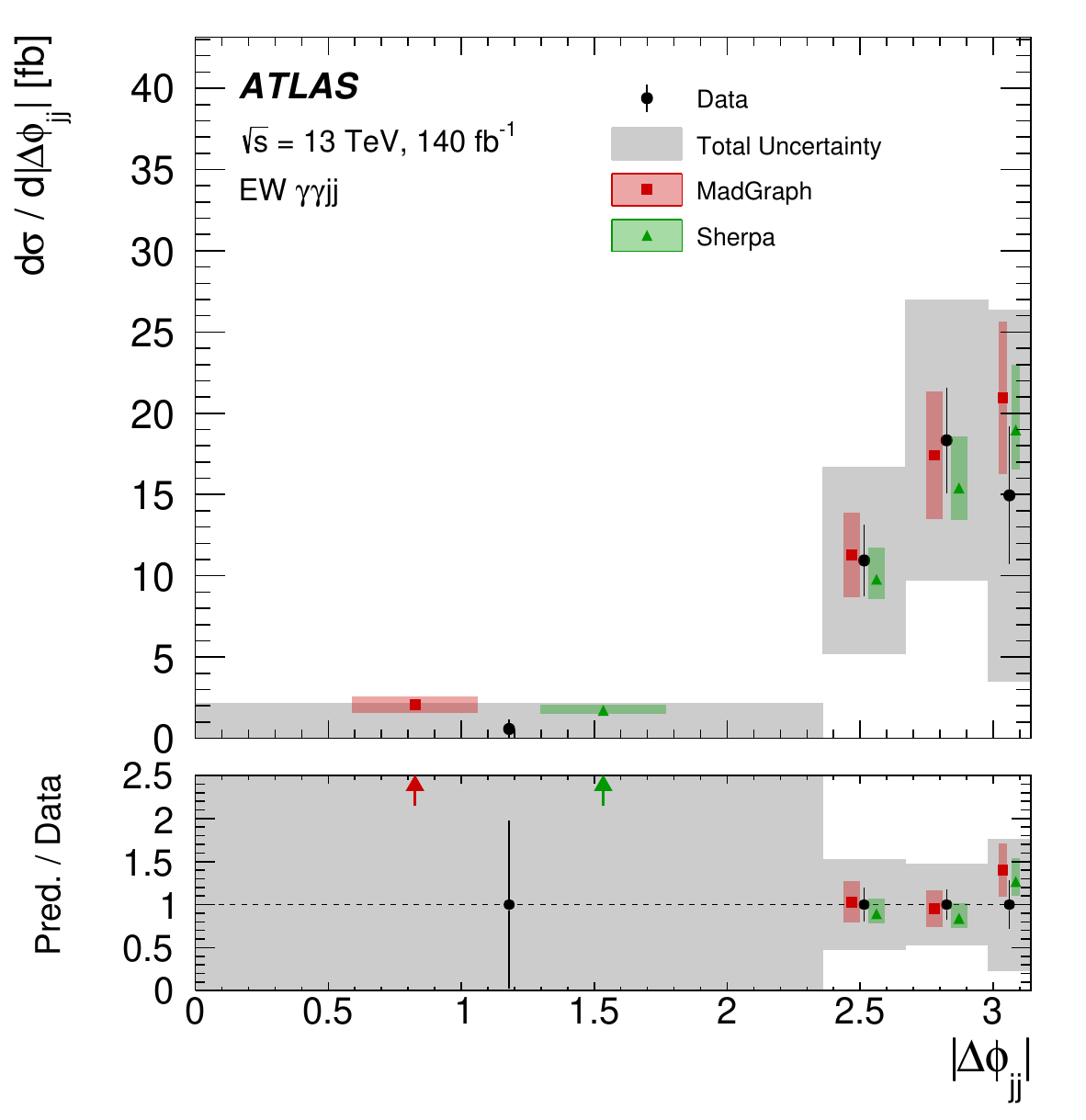}}

\caption{Differential EW-only $\gamma\gamma jj$ measurements for:
(a) $m_{jj}$,
(b) $m_{\gamma\gamma}$,
(c) $p_{\mathrm{T},\gamma_1}$,
(d) $p_{\mathrm{T},j_1}$,
(e) $p_{\mathrm{T},j_2}$, and
(f) $|\Delta\phi_{jj}|$.
The lower panel shows the ratio of the predictions to the data.
The black dots represent the data points with the statistical uncertainties, while the grey band represents the total uncertainties.
Particle-level predictions of EW-$\gamma\gamma jj$ are shown for \MGNLO interfaced with \PYTHIA and for \SHERPA. The coloured area around the predictions represents the effect of the systematic uncertainties from the renormalisation and factorisation scales.  Except for $|\Delta\phi_{jj}|$, the last bin includes overflow events. }

\label{fig:EResults1}
\end{figure}


\section{EFT interpretation}
\label{sec:EFT}
An EFT approach is used for parametrising the effects of anomalous dimension-8 gauge-bosons couplings.  In the EFT framework, new physics contributions are assumed to be at higher energy scales ($\Lambda$) than those of the SM. The SM Lagrangian ($\mathcal{L}_{SM}$) is extended with the addition of higher dimensional operators ($O_i^{(d)}$) of dimension $d>4$:

\begin{equation}
\mathcal{L}_{EFT} =\mathcal{L}_{SM} +\sum_i \frac{c_i^{(d)}}{\Lambda^{(d-4)}}O_i^{(d)},
\end{equation}

with $c_i^{(d)} $ the Wilson coefficients describing the strength of the new interactions. In this Letter only dimension-8 operators~\cite{EboliModel} are considered, since these are the only terms that modify the quartic couplings, and since dimension-6 operators relevant to triple gauge-boson couplings were already tightly constrained by analyses of diboson processes~\cite{STDM-2016-15,STDM-2017-03,STDM-2017-18,STDM-2017-24,CMS-SMP-16-017,CMS-SMP-16-018,CMS-SMP-18-004,CMS-SMP-19-001}. In the EFT basis described in Ref.~\cite{EboliModel},  the relevant operators are those containing four field-strength tensors, $T_i$,  with corresponding Wilson coefficients $ f_{T_i}$.
Events are generated separately for the SM contribution, the interference between the SM and the BSM operators, and the term quadratic in the BSM operators. Each dimension-8 operator is considered independently, ignoring effects from possible interference between different operators.  The effective differential cross-section as a function of an observable $x$ is then parameterised as:
\begin{equation}
\frac{d\sigma_{EFT}}{dx}=\frac{d\sigma_{SM}}{dx}+f_{O_i}\cdot \frac{d\sigma_{\textrm{lin}}}{dx}+f_{O_i}^2\cdot \frac{d\sigma_{\textrm{quad}}}{dx}.
\end{equation}

A maximum-likelihood fit is performed over the differential EW-only \yyjj results of $\pTyo$,  since, between all the measured observables, it provides maximum sensitivity to the Wilson coefficients.  Most of the EFT effects are concentrated in the last bin of \pTyo. The likelihood function is built with a multivariate Gaussian function,  considering the covariance matrix provided from the unfolded measurement. The SM contribution normalisation and the Wilson coefficients $f_{O_i}$ are left floating in the likelihood.
Systematic uncertainties on the BSM predictions are considered, including those related to the PDFs and scale uncertainties.  No statistically significant deviations relative to the SM are observed, and limits are set on the Wilson coefficients. Table~\ref{tab:EFT} shows the observed and expected 95\% confidence level intervals for each Wilson coefficient considered. The highest sensitivity is achieved with an observed 95\% confidence level (CL) interval of $[-0.400,\; 0.372]~\TeV^{-4}$ for $f_{T5}/\Lambda^4$, and $[-0.220,\; 0.219]~\TeV^{-4}$ for $f_{T8}/\Lambda^4$.
The largest impacts on the sensitivity to the Wilson coefficients arise from the statistical uncertainty on the non-prompt background estimate and the theoretical uncertainties in the SM contribution. The current best 95\% CL limits on the same operators are provided by the $Z(\nu\bar{\nu})\gamma jj$ ATLAS analysis~\cite{STDM-2018-59}, with intervals of $[-0.088,\;0.099]~\TeV^{-4}$ for $f_{T5}/\Lambda^4$ and $[-0.059,\;0.059]~\TeV^{-4}$ for $f_{T8}/\Lambda^4$.

\begin{table}[ht]
\caption{Expected and observed lower and upper limits at 95\% CL for the Wilson coefficients of dimension-8-operators from the fit to the transverse momentum of the leading photon $p^T_{\gamma_1}$. No unitarity constraints are applied.}
\centering
\begin{tabular}{ccc}
\hline
Coefficient & Expected Limits [~\TeV$^{-4}$] & Observed Limits [~\TeV$^{-4}$]  \\
\hline
$f_{T0}/\Lambda^4$ & $[-1.29,\; 1.22]$ & $[-1.16,\; 1.09]$ \\
$f_{T1}/\Lambda^4$ & $[-1.61,\; 1.59]$ & $[-1.45,\; 1.43]$ \\
$f_{T2}/\Lambda^4$ & $[-3.29,\; 3.15]$ & $[-2.97,\; 2.82]$ \\
$f_{T3}/\Lambda^4$ & $[-3.42,\; 3.26]$ & $[-3.08,\; 2.92]$ \\
$f_{T4}/\Lambda^4$ & $[-1.73,\; 1.61]$ & $[-1.57,\; 1.44]$ \\
$f_{T5}/\Lambda^4$ & $[-0.443,\; 0.416]$ & $[-0.400,\; 0.372]$ \\
$f_{T6}/\Lambda^4$ & $[-0.987,\; 0.972]$ & $[-0.890,\; 0.872]$ \\
$f_{T7}/\Lambda^4$ & $[-1.48,\; 1.39]$ & $[-1.34,\; 1.24]$ \\
$f_{T8}/\Lambda^4$ & $[-0.244,\; 0.244]$ & $[-0.220,\; 0.219]$ \\
$f_{T9}/\Lambda^4$ & $[-0.524,\; 0.524]$ & $[-0.471,\; 0.471]$ \\
\hline
\end{tabular}
\label{tab:EFT}
\end{table}

Limits on the Wilson coefficients are also evaluated with the `clipping` techniques~\cite{Brivio:2022pyi} as a function of a cut-off scale $\sqrt{\hat{s}} $, defined as the invariant mass of the diphoton system, by setting the EFT contribution to zero for events with $\sqrt{\hat{s}} $ greater than the cut-off scale. Unitarity constraints for the anomalous dimension-8 operators are evaluated following Ref.~\cite{unitarityconstr}. Figure~\ref{fig:t5clipping} shows the 95\% CL for the two most sensitive Wilson coefficients $f_{T5}/\Lambda^4$ and  $f_{T8}/\Lambda^4$   as a function of \myy, together with the unitarity limits.

\begin{figure}[!htbp]
\centering
\subfloat[]{\includegraphics[width=0.4\textwidth]{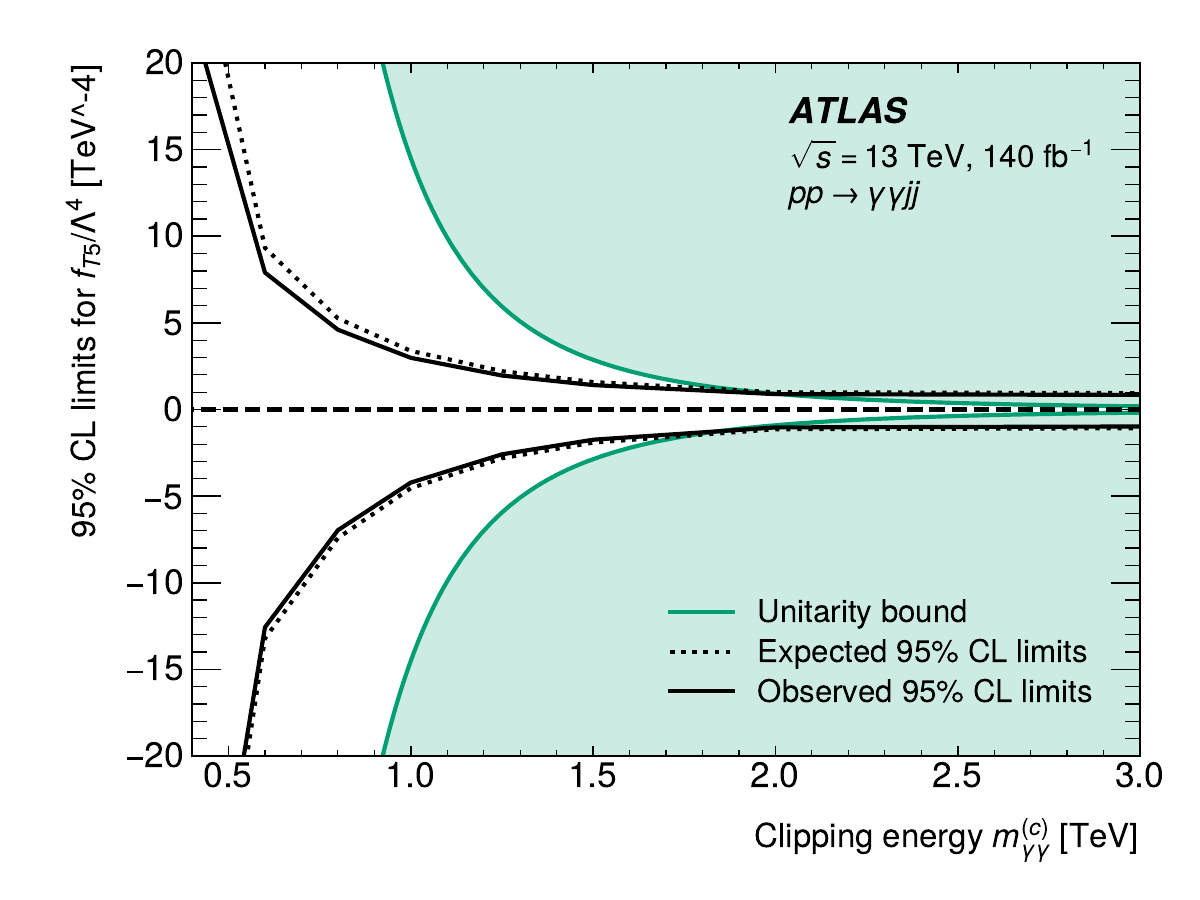}}
\subfloat[]{\includegraphics[width=0.4\textwidth]{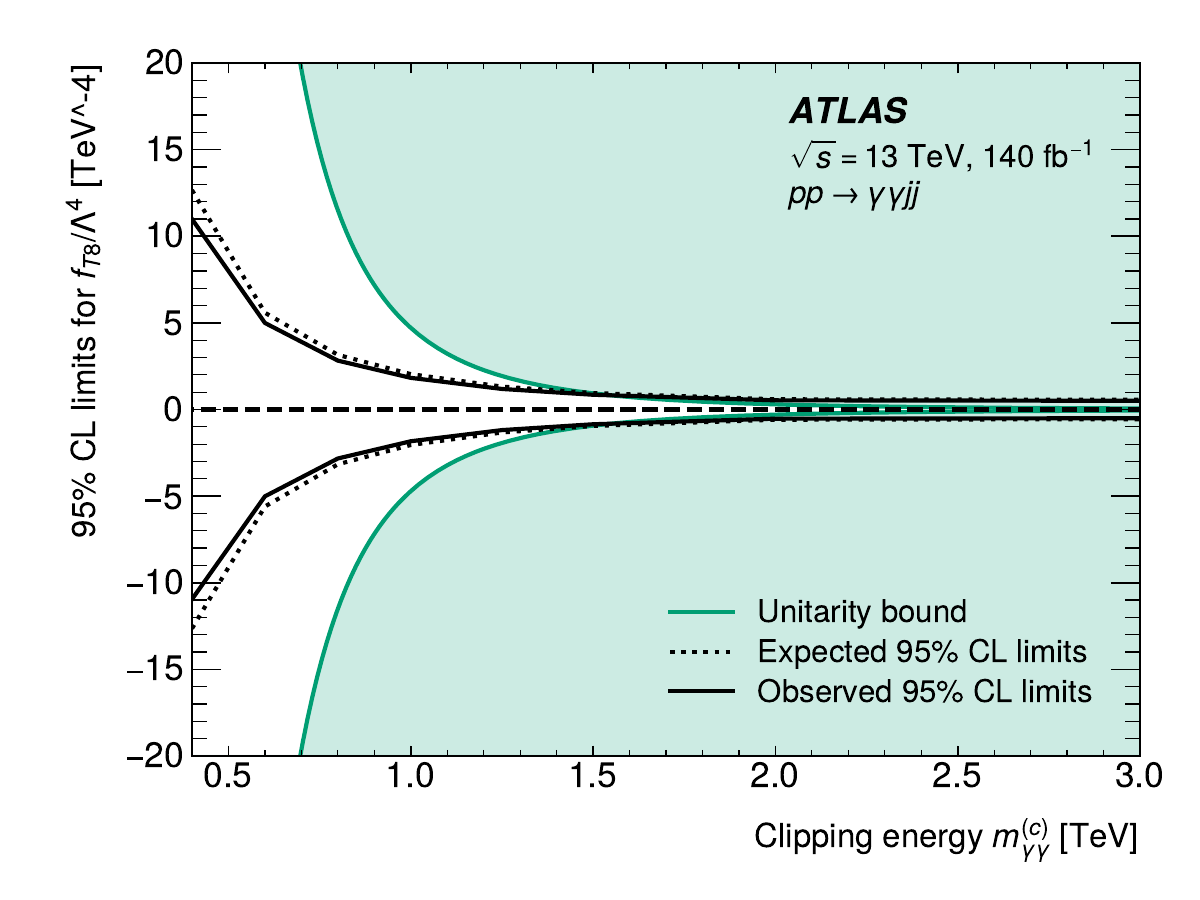}} \\
\caption{ Expected  and observed lower and upper limit for the (a) $f_{T5}$ and (b) $f_{T8}$ coefficient from the fit to the transverse momentum of the leading photon at different cutoff values ($m^{(c)}_{\gamma\gamma}$) of \myy, compared with the unitarity bounds. The green shaded area corresponds to the regions excluded by unitarity constraints.}
\label{fig:t5clipping}
\end{figure}


\section{Conclusion}
\label{sec:conclusion}

This Letter presents the observation of the EW production of \yyjj events and measurements of the EW-\yyjj and combined QCD+EW \yyjj processes. The analysis uses data collected by the ATLAS experiment at the LHC with $pp$ collisions at centre-of-mass energy of 13~\TeV for an integrated luminosity of 140~\ifb. The analysis selects two isolated photons and two high transverse momentum jets. A data-driven method is used to extract the fraction of background events coming from non-prompt photons. In the analysis, the EW-\yyjj signal is extracted from a fit to \mjj in a signal region based on the centrality of the photons ($\xi_{\gamma\gamma}$), and the number of gap-jets ($\Ngy$).
The EW-\yyjj signal strength is $\mu^{\mathrm{EW}}_{\mathrm{obs}}=0.80^{+0.18}_{-0.15}$ with an observed significance of 6.2$\sigma$, corresponding to a measured EW-\yyjj cross--section of $\sigma^{\mathrm{EW}}_{\mathrm{obs}}= 13.8^{+3.0}_{-2.6}\, \mathrm{fb}$, consistent with the SM predictions. The measured QCD+EW \yyjj cross-section is $\sigma_{\mathrm{obs}}^{\mathrm{QCD+EW}}=54.3^{+4.6}_{-4.8}\, \mathrm{fb}$.
Differential measurements are performed of the EW-\yyjj process and for the combined QCD+EW \yyjj process. Measurements are performed as a function of the diphoton and dijet invariant masses and transverse momenta (\mjj, \myy, $p_{T,jj}$, $p_{T,\gamma\gamma}$), of the photons and jets transverse momenta ($p_{T,\gamma_{1(2)}}$, $p_{T,j_{1(2)}}$), of several angular  distributions ($\Delta y_{jj}$,  $|\Delta\phi_{jj}|$,  $\Delta\phi_{jj}^{\text{sign}}$) and of  $\xi_{\gamma\gamma}$ and  $\Ngy$ (only for the  QCD+EW \yyjj ). Uncertainties vary depending on the observable and the bin, from 10\% to 40\% for the measurement of QCD+EW \yyjj, while they are above 100\% in a few bins in the EW-only \yyjj measurement, and good agreement is observed with predictions.
Additionally, the measurement of leading photon momentum ($\pTyo$) is interpreted in an EFT framework. No deviations relative to the SM are observed, and limits on the Wilson coefficients of the anomalous dimension-8 operators are set. In particular, the most stringent limits are set on $f_{T5}/\Lambda^4$ and $f_{T8}/\Lambda^4$ with a corresponding observed 95\% CL limit of $[-0.400,\; 0.372]~\TeV^{-4}$ and $[-0.220,\; 0.219]~\TeV^{-4}$.


%
%
%
%
%
\section*{Acknowledgements}
%

%

%
%

%
%

We thank CERN for the very successful operation of the LHC and its injectors, as well as the support staff at
CERN and at our institutions worldwide without whom ATLAS could not be operated efficiently.

The crucial computing support from all WLCG partners is acknowledged gratefully, in particular from CERN, the ATLAS Tier-1 facilities at TRIUMF/SFU (Canada), NDGF (Denmark, Norway, Sweden), CC-IN2P3 (France), KIT/GridKA (Germany), INFN-CNAF (Italy), NL-T1 (Netherlands), PIC (Spain), RAL (UK) and BNL (USA), the Tier-2 facilities worldwide and large non-WLCG resource providers. Major contributors of computing resources are listed in Ref.~\cite{ATL-SOFT-PUB-2026-001}.

We gratefully acknowledge the support of ANPCyT, Argentina; YerPhI, Armenia; ARC, Australia; BMWFW and FWF, Austria; ANAS, Azerbaijan; CNPq and FAPESP, Brazil; NSERC, NRC and CFI, Canada; CERN; ANID, Chile; CAS, MOST, NSFC and FRFCU, China; Minciencias, Colombia; MEYS CR, Czech Republic; DNRF and DNSRC, Denmark; IN2P3-CNRS and CEA-DRF/IRFU, France; SRNSFG, Georgia; BMFTR, HGF and MPG, Germany; GSRI, Greece; RGC and Hong Kong SAR, China; ICHEP and Academy of Sciences and Humanities, Israel; INFN, Italy; MEXT and JSPS, Japan; CNRST, Morocco; NWO, Netherlands; RCN, Norway; MNiSW, Poland; FCT, Portugal; MNE/IFA, Romania; MSTDI, Serbia; MSSR, Slovakia; ARIS and MVZI, Slovenia; DSI/NRF, South Africa; MICIU/AEI, Spain; SRC and Wallenberg Foundation, Sweden; SERI, SNSF and Cantons of Bern and Geneva, Switzerland; NSTC, Taipei; TENMAK, T\"urkiye; STFC/UKRI, United Kingdom; DOE and NSF, United States of America.

Individual groups and members have received support from BCKDF, CANARIE, CRC and DRAC, Canada; CERN-CZ, FORTE and PRIMUS, Czech Republic; COST, ERC, ERDF, Horizon 2020 and Marie Sk{\l}odowska-Curie Actions, European Union; Investissements d'Avenir Labex, Investissements d'Avenir Idex and ANR, France; DFG and AvH Foundation, Germany; Herakleitos, Thales and Aristeia programmes co-financed by EU-ESF and the Greek NSRF, Greece; BSF-NSF and MINERVA, Israel; NCN and NAWA, Poland; La Caixa Banking Foundation, CERCA and AGAUR programs from Generalitat de Catalunya and PROMETEO and GenT Programmes Generalitat Valenciana, Spain; G\"{o}ran Gustafssons Stiftelse, Sweden; The Royal Society and Leverhulme Trust, United Kingdom; Eric and Wendy Schmidt Fund for Strategic Innovation, United States of America.

In addition, individual members wish to acknowledge support from Chile: Agencia Nacional de Investigaci\'on y Desarrollo (ANID FONDECYT reg. 1230987, FONDECYT 1230812, FONDECYT 1240864, Fondecyt 3240661, Fondecyt Regular 1240721); China: Fundamental Research Funds for the Central Universities (010-63263105), Chinese Ministry of Science and Technology (MOST-2023YFA1605700, MOST-2023YFA1609300), National Natural Science Foundation of China (NSFC 12275265, NSFC-W2543005); Czech Republic: Czech Science Foundation (GACR - 24-11373S), Ministry of Education Youth and Sports (ERC-CZ-LL2327, FORTE CZ.02.01.01/00/22\_008/0004632); EU: H2020 European Research Council (ERC - 101002463); European Union: European Research Council (BARD No. 101116429, ERC - 101219398, ERC 101089007), European Regional Development Fund (HE COFUND GA No.101081355, ERDF), Marie Sklodowska-Curie Actions (GAP-101168829); France: Agence Nationale de la Recherche (ANR-22-EDIR-0002, ANR-24-CE31-0504-01); Germany: Deutsche Forschungsgemeinschaft (DFG - 469666862); China: Research Grants Council (GRF); Italy: Istituto Nazionale di Fisica Nucleare (LHC-MIUR - 28003/2025), Ministero dell'Università e della Ricerca (NextGenEU I53D23000820006 M4C2.1.1, SOE2024\_0000023); Japan: Japan Society for the Promotion of Science (JSPS KAKENHI  JP25H0063, JSPS KAKENHI JP22H04944, JSPS KAKENHI JP24K23939, JSPS KAKENHI JP24KK0251, JSPS KAKENHI JP25H00650, JSPS KAKENHI JP25H01291, JSPS KAKENHI JP25K01011, JSPS KAKENHI JP25K01023, JSPS KAKENHI JP25KK0047); Poland: Polish National Science Centre (NCN 2021/42/E/ST2/00350, NCN OPUS 2023/51/B/ST2/02507, NCN OPUS nr 2022/47/B/ST2/03059, NCN UMO-2019/34/E/ST2/00393, UMO-2022/47/O/ST2/00148, UMO-2023/49/B/ST2/04085, UMO-2023/51/B/ST2/00920, UMO-2024/53/N/ST2/00869); Spain: Agència de Gestió d'Ajuts Universitaris i de Recerca. (AGAUR - 2023 BP 00141), Ministry of Science and Innovation (RYC2019-028510-I, RYC2020-030254-I, RYC2021-031273-I, RYC2022-038164-I), Ministerio de Ciencia, Innovación y Universidades/Agencia Estatal de Investigaci\'on (EU NextGenerationEU (PRTR-C17.I1), PID2022-142604OB-C22); Sweden: Carl Trygger Foundation (Carl Trygger Foundation CTS 22:2312), Swedish Research Council (Swedish Research Council 2023-04654, VR 2021-03651, VR 2022-03845, VR 2022-04683, VR 2023-03403, VR 2024-05451, VR 2025-05940), Knut and Alice Wallenberg Foundation (KAW 2023.0366); United Kingdom: The Binks Trust, Royal Society (NIF-R1-231091); United States of America: U.S. Department of Energy (ECA DE-AC02-76SF00515), John Templeton Foundation (John Templeton Foundation 63206), Neubauer Family Foundation.

%
%


%
\clearpage
\appendix
\printbibliography
\clearpage
\input{atlas_authlist}
%
\clearpage

\end{document}

%% file: atlas_authlist.tex
 
\begin{flushleft}
\hypersetup{urlcolor=black}
{\Large The ATLAS Collaboration}

\bigskip

\AtlasOrcid[0000-0002-6665-4934]{G.~Aad}$^\textrm{\scriptsize 102}$,
\AtlasOrcid[0000-0001-7616-1554]{E.~Aakvaag}$^\textrm{\scriptsize 17}$,
\AtlasOrcid[0000-0002-5888-2734]{B.~Abbott}$^\textrm{\scriptsize 121}$,
\AtlasOrcid[0000-0002-0287-5869]{S.~Abdelhameed}$^\textrm{\scriptsize 83b}$,
\AtlasOrcid[0000-0002-1002-1652]{K.~Abeling}$^\textrm{\scriptsize 54}$,
\AtlasOrcid[0000-0001-5763-2760]{N.J.~Abicht}$^\textrm{\scriptsize 48}$,
\AtlasOrcid[0000-0002-8496-9294]{S.H.~Abidi}$^\textrm{\scriptsize 30}$,
\AtlasOrcid[0009-0003-6578-220X]{M.~Aboelela}$^\textrm{\scriptsize 44}$,
\AtlasOrcid[0000-0002-9987-2292]{A.~Aboulhorma}$^\textrm{\scriptsize 36e}$,
\AtlasOrcid[0000-0001-5329-6640]{H.~Abramowicz}$^\textrm{\scriptsize 154}$
% ...
\bigskip
\\

% ...
$^{17}$Department for Physics and Technology, University of Bergen, Bergen; Norway.\\
% ...
$^{30}$Physics Department, Brookhaven National Laboratory, Upton NY; United States of America.\\
% ...
$^{36}$$^{(a)}$Facult\'e des Sciences Ain Chock, Universit\'e Hassan II de Casablanca;$^{(b)}$Facult\'{e} des Sciences, Universit\'{e} Ibn-Tofail, K\'{e}nitra;$^{(c)}$Facult\'e des Sciences Semlalia, Universit\'e Cadi Ayyad, LPHEA-Marrakech;$^{(d)}$LPMR, Facult\'e des Sciences, Universit\'e Mohamed Premier, Oujda;$^{(e)}$Facult\'e des sciences, Universit\'e Mohammed V, Rabat;$^{(f)}$Institute of Applied Physics, Mohammed VI Polytechnic University, Ben Guerir; Morocco.\\
% ...
$^{44}$Physics Department, Southern Methodist University, Dallas TX; United States of America.\\
% ...
$^{48}$Fakult\"{a}t Physik, Technische Universit{\"a}t Dortmund, Dortmund; Germany.\\
% ...
$^{54}$II. Physikalisches Institut, Georg-August-Universit\"{a}t G\"ottingen, G\"ottingen; Germany.\\
% ...
$^{83}$$^{(a)}$Khalifa University of Science and Technology, Abu Dhabi;$^{(b)}$New York University Abu Dhabi, Abu Dhabi;$^{(c)}$United Arab Emirates University, Al Ain;$^{(d)}$University of Sharjah, Sharjah; United Arab Emirates.\\
% ...
$^{102}$CPPM, Aix-Marseille Universit\'e, CNRS/IN2P3, Marseille; France.\\
% ...
$^{121}$Homer L. Dodge Department of Physics and Astronomy, University of Oklahoma, Norman OK; United States of America.\\
% ...
$^{154}$Raymond and Beverly Sackler School of Physics and Astronomy, Tel Aviv University, Tel Aviv; Israel.\\
% ...
$^{a}$ Also at Affiliated with an institute formerly covered by a cooperation agreement with CERN.\\
$^{b}$ Also at An-Najah National University, Nablus; Palestine.\\
$^{c}$ Also at Borough of Manhattan Community College, City University of New York, New York NY; United States of America.\\
$^{d}$ Also at Center for Interdisciplinary Research and Innovation (CIRI-AUTH), Thessaloniki; Greece.\\
$^{e}$ Also at Centre of Physics of the Universities of Minho and Porto (CF-UM-UP); Portugal.\\
$^{f}$ Also at CERN, Geneva; Switzerland.\\
% ...

\end{flushleft}
